\documentclass[preprint,amssymb,aip,jap,numerical,superscriptaddress]{revtex4-1}

\usepackage{amsmath}    
\usepackage{graphicx}   
\usepackage{verbatim}   
\usepackage{color}      
\usepackage{subfigure}	
\usepackage{hyperref}   
\usepackage{mathtools}

\newcommand{\eqs}{\begin{eqnarray*}}
\newcommand{\eqf}{\end{eqnarray*}}
\newcommand{\lef}{\left(}
\newcommand{\rig}{\right)}
\newcommand{\mas}{\begin{array}}
\newcommand{\maf}{\end{array}}
\newcommand{\deriv}{\textnormal{d}}

\begin{document}

\title{Modeling surface roughness scattering in metallic nanowires}
\author{Kristof Moors}
\email[E-mail me at: ]{kristof@itf.fys.kuleuven.be}
\affiliation{KU Leuven, Institute for Theoretical Physics, Celestijnenlaan 200D, B-3001 Leuven, Belgium}
\affiliation{Imec, Kapeldreef 75, B-3001 Leuven, Belgium}
\author{Bart Sor\'ee}
\affiliation{Imec, Kapeldreef 75, B-3001 Leuven, Belgium}
\affiliation{University of Antwerp, Physics Department, Groenenborgerlaan 171, B-2020 Antwerpen, Belgium}
\affiliation{KU Leuven, Electrical Engineering (ESAT) Department, Kasteelpark Arenberg 10, B-3001 Leuven, Belgium}
\author{Wim Magnus}
\affiliation{Imec, Kapeldreef 75, B-3001 Leuven, Belgium}
\affiliation{University of Antwerp, Physics Department, Groenenborgerlaan 171, B-2020 Antwerpen, Belgium}

\date{\today}



\begin{abstract}
Ando's model provides a rigorous quantum-mechanical framework for electron-surface roughness scattering, based on the detailed roughness structure. We apply this method to metallic nanowires and improve the model introducing surface roughness distribution functions on a finite domain with analytical expressions for the average surface roughness matrix elements. This approach is valid for any roughness size and extends beyond the commonly used Prange-Nee approximation. The resistivity scaling is obtained from the self-consistent relaxation time solution of the Boltzmann transport equation and is compared to Prange-Nee's approach and other known methods. The results show that a substantial drop in resistivity can be obtained for certain diameters by achieving a large momentum gap between Fermi level states with positive and negative momentum in the transport direction.
\end{abstract}

\maketitle

\section{Introduction}
Modeling surface roughness (SR) scattering of electrons in metallic nanowires is quite challenging and often requires drastic approximations. Early attempts by Fuchs,\cite{fuchs1938conductivity} and later Sondheimer,\cite{sondheimer1952mean} introduced a specularity parameter $p$ (also known as the Fuchs parameter) that represents a probability for diffusive scattering at the boundary with the electron losing forward momentum. While predicting the experimentally observed 1/width behavior of the resistivity in thin metallic films or wires,\cite{durkan2000size,steinhogl2002size,steinlesberger2002electrical,guillaumond2003analysis,wu2004influence,steinhogl2004comprehensive,zhang2007analysis,josell2009size,graham2010resistivity,chawla2011electron} this assumption artificially imposes a boundary condition on the Boltzmann distribution function, neglecting the quantum-mechanical nature of the scattering events, the impact of confinement and the detailed roughness profile. It is therefore expected that the Fuchs-Sondheimer model, or the extension by Mayadas and Shatzkes to include grain boundary scattering,\cite{mayadas1970electrical} fails when the diameter of the metallic wire is reduced and becomes comparable with the roughness size or enters the regime where quantum-mechanical effects such as energy quantization become important.

A predictive model for SR scattering is very important, because it is commonly used to extract information on the relative contribution to the overall resistivity, also coming from other scattering mechanisms such as phonons or grain boundaries. It is commonly stated that SR scattering is one of the dominant scattering mechanisms in metallic nanowires and therefore an important bottleneck for further downscaling of semiconductor devices, where they serve as interconnects. One should be careful distinguishing between resistivity contributions relying on the Fuchs-Sondheimer or Mayadas-Shatzkes model because they could be invalid for the dimensions of the wires in the experimental setup. The relative contribution of SR scattering to the resistivity could be properly confirmed if the results of simulations and experimental data would appear to be in agreement for the resistivity values and scaling with width (or diameter), without the need of non-physical fitting parameters such as the Fuchs parameter. A model that only depends on the details of SR (or its average properties) is therefore needed.

Being a well-known approach developed to this end, Ando's model employs Fermi's golden rule to calculate the SR scattering probability, thereby relying merely on two statistical properties, the SR standard deviation and the SR
correlation length.\cite{ando1982electronic}. A commonly used approximation is the Prange-Nee approximation,\cite{prange1968quantum} replacing the overlap of wave functions in the confinement directions by the value of the wave functions at the boundary times the barrier height in the limit of an infinite barrier. Recently, the resistivity was obtained for narrow metallic nanowires (width and height ranging from 1 to 6~nm), using the Prange-Nee approximation in Ando's model. It was shown that the resistivity due to SR can have very different values and scaling behavior for different values of SR standard deviation and correlation length.\cite{moors2014resistivity}

A disadvantage of the Prange-Nee approximation is that it takes the infinite barrier limit of an expansion for small roughness sizes, which often falls short to determine the wave function overlap, especially in the case of large roughness sizes, highly oscillating wave functions or low barrier heights. For metallic nanowires one should definitely be careful, as there are many subbands and, hence, high wave vector values in the confinement direction are generally present. Quite recently, a new roughness model was proposed to improve the computation of the wave function overlap in thin-body fully depleted SOI device simulations, using a bivariate normal distribution function for the roughness statistics.\cite{lizzit2014new} This method requires numerical integration for the large amount of subbands in metallic nanowires drastically slows down the simulation. Therefore we propose two variants of a novel approach, based on distribution functions on a finite domain. In the regime of wire dimensions that can be simulated with all methods mentioned above, we will make the comparison.

The paper is structured as follows. In section \ref{sectionSRS} the SR formulation and statistics are briefly summarized. In section \ref{sectionFGR} this formulation is used to retrieve the SR matrix elements, adopting the Prange-Nee approximation, the bivariate normal distribution function and the two variants of the new approach, based on finite domain distribution functions. We also briefly discuss how the SR matrix elements are used in the Boltzmann equation and the conductivity calculation, but this can be found in more detail in Moors et al.\cite{moors2014resistivity} Results are given and discussed in section \ref{sectionResultsDiscussion} for a single subband toy model and more realistically sized metallic wires, followed by a conclusion in section \ref{sectionConclusion}.

\section{Surface roughness}
\label{sectionSRS}
Consider an ideal wire with length $L_z$ and rectangular cross section (width $L_x$ and height $L_y$), represented by a rectangular finite potential well with barrier height $U$. The ideal wire Hamiltonian $H_0$ is then given by:
\begin{align*}
H_0 \lef \mathbf{r} \rig &=
\left\{
\begin{matrix}
\; -\frac{\hbar^2}{2 m_e} \nabla^2, \qquad \quad \textnormal{if } 0 \leq x/y \leq L_{x/y}, \; |z| \leq L_z/2 \\
\mkern-35mu -\frac{\hbar^2}{2 m_e} \nabla^2 + U, \quad \textnormal{else} \qquad \quad \qquad \qquad \qquad
\end{matrix} \right. ,
\end{align*}
with an effective mass $m_e$ assumed to be appropriate for the conduction electrons in the metallic nanowire under consideration. Periodic boundary conditions in the transport direction $z$ are also assumed, neglecting any effect of the contacts.
SR is modeled by SR functions that describe the deviation from the smooth, ideal boundary surface for each of the four boundaries: $S^{x=0}(y,z)$, $S^{x=L_x}(y,z)$, $S^{y=0}(x,z)$ and $S^{y=L_y}(x,z)$. Typically one supposes Gaussian or exponential autocorrelation functions for the SR functions and in this paper we consider them to be Gaussian,
\begin{align}
\label{gaussianCorr}
\left< S^{x=0}(y,z) S^{x=0}(\tilde{y},\tilde{z}) \right> &= \Delta^2 e^{-[(y - \tilde{y})^2 + (z-\tilde{z})^2]/(\Lambda^2/2) },
\end{align}
with standard deviation $\Delta$ and correlation length $\Lambda$. We only consider SR around the ideal wire for each boundary separately thus neglecting edge effects of the autocorrelation functions at the intersection of two boundaries. The autocorrelation functions will be used in the following section to calculate all the SR matrix elements.

\section{Transition probability matrix elements}
\label{sectionFGR}
\begin{figure}[tb]
\centering
\includegraphics[width=0.5\linewidth]{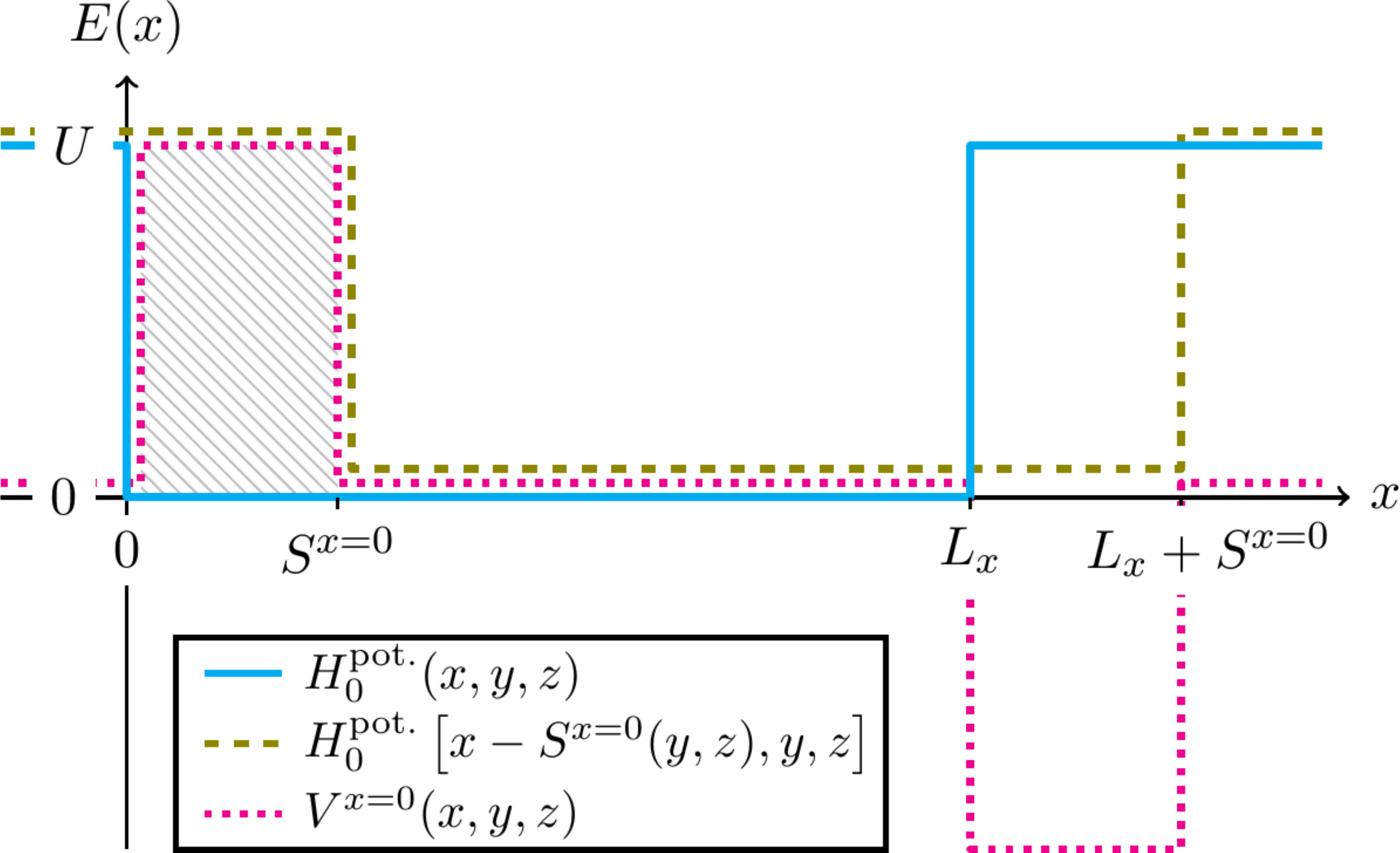}
\caption{The potential energy part of the ideal Hamiltonian, $H_0^\textnormal{pot.}(\mathbf{r})$, is shown (continuous line) as a function of $x$ for a fixed $y$ and $z$, together with the Hamiltonian shifted by $S^\textnormal{x=0}(y,z)$ (dashed line) and their difference $V^{x=0}(x,y,z)$ (dotted line). To obtain the potential for SR of the $x=0$ boundary, only the part of $V^{x=0}$ close to $x=0$ should be considered. This part is shown hatched on the figure.}
\label{SRPotential}
\end{figure}
The probability for an electron to scatter from an initial state $\mid i \rangle$ to a final state $\mid f \rangle$ due to a perturbation Hamiltonian $V$ is given by
\begin{align}
\label{FGR}
P\lef \mid i \rangle \rightarrow \mid f \rangle \rig &= \frac{2\pi}{\hbar} \left| \langle i \mid V \mid f \rangle \right|^2 \delta\lef E_i - E_f \rig,
\end{align}
according to Fermi's golden rule with $E_i$ and $E_f$ the energy of initial and final state respectively. The scattering probabilities enter the collision term of the Boltzmann transport equation (BTE) through the relaxation time approximation, such that the following equations have to be solved to obtain the relaxation times:
\begin{align}
\label{RTA}
\frac{1}{\tau_i} = \frac{m_e L_z}{\hbar^3} \sum_{f} \frac{\left| \langle i \mid V \mid f \rangle \right|^2}{| k_f^z |} \lef 1 - \frac{k^z_f \tau_f }{k^z_i \tau_i} \rig,
\end{align}
where we assumed running wave solutions along the transport direction with wave vectors $k^z_i$ and $k^z_f$ for the initial and final state. The relaxation times can be calculated self-consistently, or by applying commonly used approximations for the ratio of relaxation times on the RHS. We only consider states at the Fermi level because SR scattering is considered to be elastic, while room temperature $k_\textnormal{\tiny B}T$ is very small compared to the Fermi level in metals,\footnote{We refer here to the energy difference between the highest occupied single-electron state in the ground state and the bottom of the potential well for the conduction electrons.} hence the deviation from the step function in the $T=0$ Fermi-Dirac distribution is negligible. Using the nanowire model introduced in section~\ref{sectionSRS}, a matrix element is in general given by:
\begin{align}
\label{matrixElement}
\langle i \mid V \mid f \rangle &= \frac{1}{L_z}\int \mkern-3mu \deriv \mathbf{r} \; \psi^*_{i} (x,y) e^{-ik^z_i z} \left[ H(\mathbf{r}) - H_0(\mathbf{r}) \right] \psi_{f} (x,y) e^{ik^z_f z},
\end{align}
with $\psi(x,y)\equiv \psi(x) \psi(y)$ and $\psi(x/y)$ the finite potential well wave functions (the argument indicates the corresponding transversal direction). For SR at the $x=0$ surface the Hamiltonian $H$ can be written as the unperturbed Hamiltonian shifted by the appropriate SR function:
\begin{align*}
H^{x=0}\lef x,y,z \rig = H_0 \left[ x - S^{x=0}(y,z), y, z \right].
\end{align*}
In this way the perturbation $V^{x=0} \equiv H^{x=0} - H_0$ can easily be obtained, as can be seen in Fig.~\ref{SRPotential}. The shift also acts around $x=L_x$, which should not influence the $x=0$ SR matrix element. Therefore the potential is always considered to be zero when $|x| > \left|S^{x=0}(y,z)\right|$ while $|S^{x=0}(y,z)|$ is assumed to be considerably smaller than $L_x$ for all $y$ and $z$ to exclude the unphysical situation in which the rough surface crosses the opposite wire boundary. The expressions for the SR matrix elements coming from the other wire boundaries are similarly obtained.

\subsection{Prange-Nee approximation}
Instead of solving the integral in Eq.~\ref{matrixElement}, the Prange-Nee approximation expands the integrand up to first order of the SR function $S^{x=0}(y,z)$ while taking the infinite barrier limit of the first order term
.\cite{prange1968quantum} For the $x=0$ surface, the expansion up to first order of the SR function yields:
\begin{align}
\label{matrixElementInfiniteWell}
& \langle i \mid V^{x=0} \mid f \rangle \approx \frac{U \left. \psi^*_i(x) \psi_f(x) \right|_{x=0}}{L_z} \int\limits_0^{L_y} \mkern-3mu \deriv y \mkern-8mu \int\limits_{-L_z/2}^{+L_z/2} \mkern-12mu \deriv z \; S^{x=0}(y,z) \psi^*_i\lef y \rig \psi_f\lef y \rig e^{-i (k^z_i - k^z_f) z}.
\end{align}
The value of $U \psi^*_i (x) \psi_f (x)$ at $x=0$ depends on the height of the potential well and the corresponding values of the wave function at the boundary. The limit for $U\rightarrow+\infty$ gives $2\sqrt{E^x_i E^x_f}/L_x$ with $E_{i/f}^x$ the energy of the initial/final state coming from the $x$-direction, which can also be written as $(\hbar^2/2m_e) \left. (\deriv/\deriv x) \psi^*_i (x) (\deriv/\deriv x) \psi_f (x) \right|_{x=0}$. To obtain an expression for the scattering probability Eq.~\ref{FGR}, the result of Eq.~\ref{matrixElementInfiniteWell} should be multiplied by its complex conjugate, making it possible to insert the correlation function of Eq.~\ref{gaussianCorr} and calculating the scattering probability analytically:\cite{moors2014resistivity}
\begin{align}
\label{PrangeNee}
\left< \left| \langle i \mid V^{x=0} \mid f \rangle \right|^2 \right> & \approx 4 \frac{E^x_i E^x_f}{L_x^2} \frac{\Delta^2}{L_z^2} \int\limits_0^{L_y} \mkern-3mu \deriv y \mkern-10mu \int\limits_{-L_z/2}^{+L_z/2} \mkern-12mu \deriv z \int\limits_0^{L_y} \mkern-3mu \deriv \tilde{y} \mkern-10mu \int\limits_{-L_z/2}^{+L_z/2} \mkern-12mu \deriv \tilde{z} \; \psi^*_i (y) \psi_f (y) \psi_f^* (\tilde{y}) \psi_i (\tilde{y}) \\ \notag
& \quad \qquad \qquad \qquad \qquad \qquad \qquad \times e^{ -i (k^z_i - k^z_f) (z - \tilde{z}) - [(y - \tilde{y})^2 + (z-\tilde{z})^2]/(\Lambda^2/2)},
\end{align}
Assuming infinite barrier height around the wire, the wave functions of the confined directions are given by $\sqrt{2/L_x}\sin\lef n_x \pi x / L_x \rig$ with positive integer subband index $n_x$ (analog for $y$-direction). The length-dependent part in the limit $L_z \gg \Lambda$ gives:
\begin{align}
\label{deltaKzCrit}
\sqrt{\frac{\pi}{2}} \frac{\Lambda}{L_z} e^{- (k^z_i-k^z_f)^2 \Lambda^2 / 8},
\end{align}
while the remaining integration over $y,\tilde{y}$ gives rise to an additional form factor $F^y_{i,f}$, such that:
\begin{align*}
\left< \left| \langle i \mid V^{x=0} \mid f \rangle \right|^2 \right> &\approx 4 \sqrt{\frac{\pi}{2}} \frac{E^x_i E^x_f}{L_x^2} \frac{\Delta^2 \Lambda}{L_z} F^y_{i,f} e^{- (k^z_i-k^z_f)^2 \Lambda^2 / 8}, \\ \notag
F^y_{i,f} &\equiv \int\limits_0^{L_y} \mkern-3mu \deriv y \int\limits_0^{L_y} \mkern-3mu \deriv \tilde{y} \; \psi^*_i (y) \psi_f (y) \psi_f^* (\tilde{y}) \psi_i (\tilde{y}) e^{ - (y - \tilde{y})^2/(\Lambda^2/2)}.
\end{align*}
The scattering probability is highly suppressed when $\left| k^z_i - k^z_f\right| > 2\sqrt{2}/\Lambda.$ With this observation we can define a critical momentum gap:
\begin{align}
\label{critMomentumGapPN}
\Delta k^z_\textnormal{\tiny crit. (PN)} \equiv 2\sqrt{2}/\Lambda.
\end{align}
The diminishing scattering probability for large $\Delta k^z$ is a crucial observation also for the improved solution of the SR matrix elements.

\subsection{Improved solution}
\label{correctSolutionSection}
\begin{figure}[tb]\begin{center}
\subfigure[\ $S^{x=0}=-0.5$~nm]{
\includegraphics[width=0.4\linewidth]{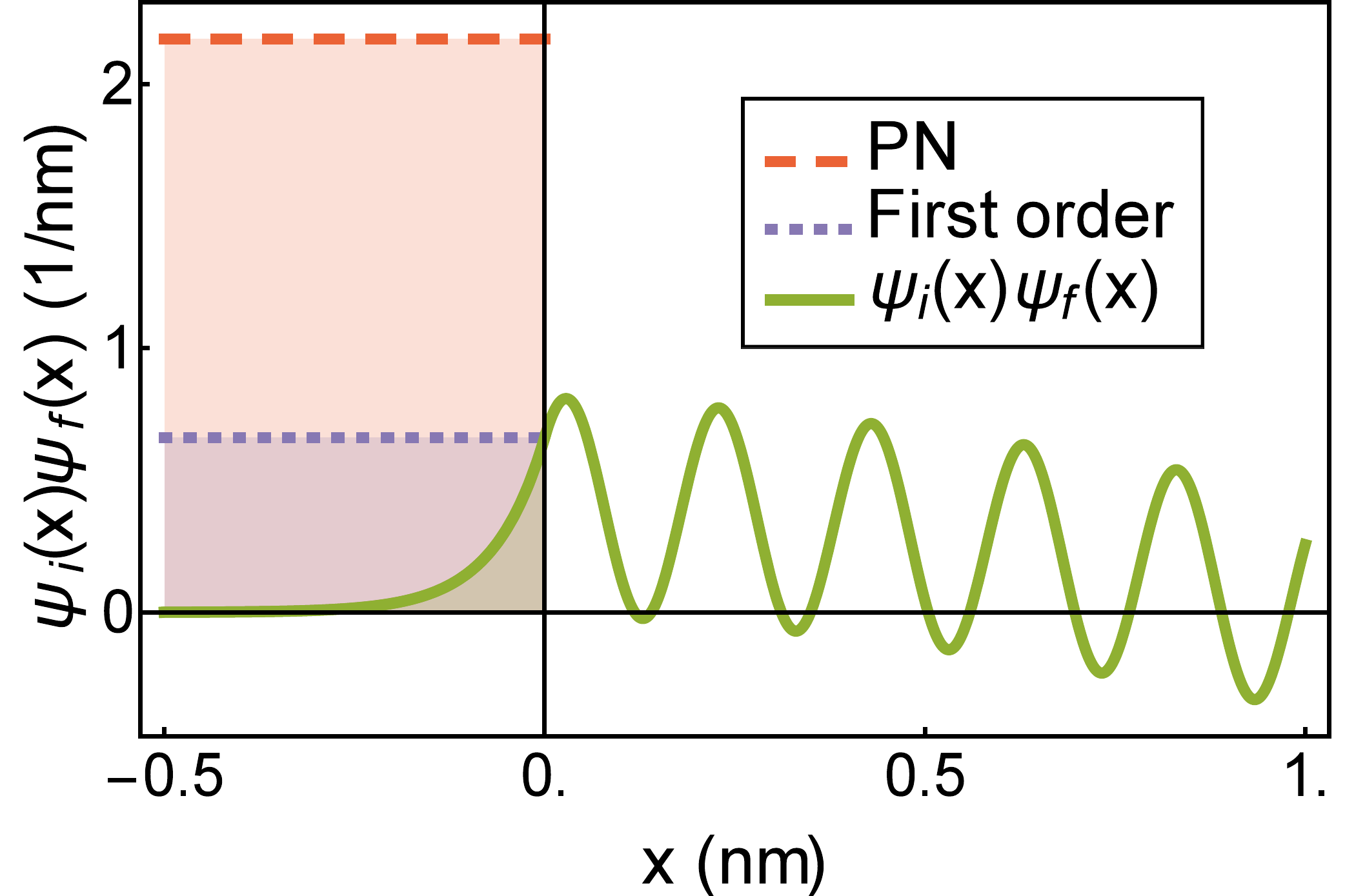}}
\subfigure[\ $S^{x=0}=1$~nm]{
\includegraphics[width=0.4\linewidth]{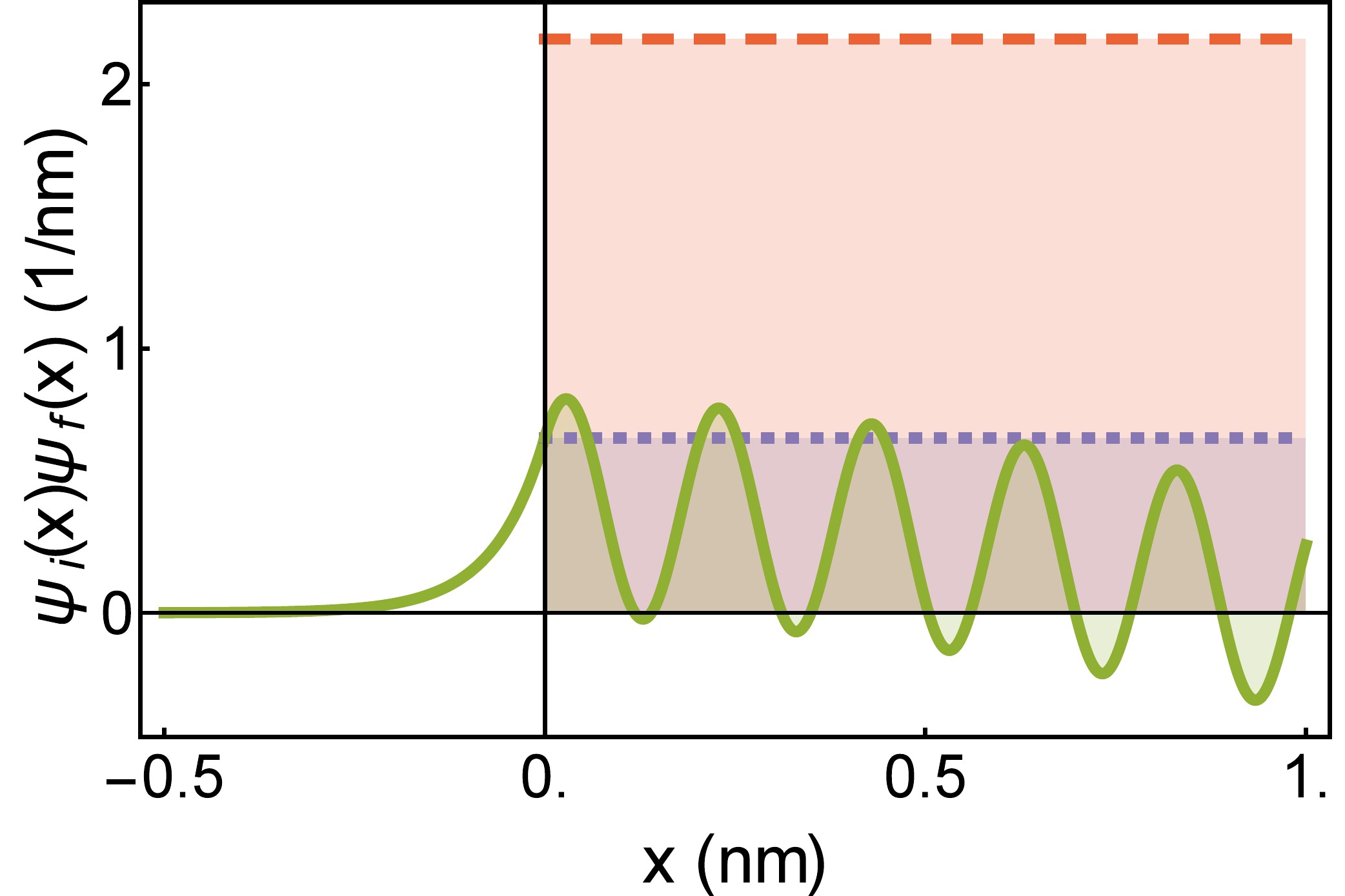}}
\end{center}
\caption{The product of wave functions with the highest and next to highest $k$ value is shown ($L_x \approx 2.15$~nm) as the green curve, with the integral of Eq.~\ref{surfaceIntegralPart} from $x=0$ up to (a) $S^{x=0}=-0.5$~nm (b) $S^{x=0}=1$~nm being the green shaded area under the curve. The Prange-Nee approximation replaces the wave function product by the red dashed line, whereas the first order approximation replaces it by the blue dotted line, such that the area under the wave function product is approximated respectively by the green or blue shaded rectangular area.}
\label{WFInt}
\end{figure}

The shortcomings of the Prange-Nee approximation become clear when we write the SR matrix element differently:
\begin{align}
\label{exactMatrixElement}
&\langle i \mid V^{x=0} \mid f \rangle = \frac{U}{L_z} \int\limits_{0}^{L_y} \mkern-3mu \deriv y \mkern-8mu \int\limits_{-L_z/2}^{+L_z/2} \mkern-12mu \deriv z \; \psi^*_i (y) \psi_f (y) e^{-i (k^z_i - k^z_f) z} \mkern-20mu \int\limits_{0}^{S^{x=0}(y,z)} \mkern-22mu \deriv x \; \psi^*_i (x) \psi_f (x).
\end{align}
The resulting expression in Eq.~\ref{exactMatrixElement} contains the integral (surface integral SI):
\begin{align}
\label{surfaceIntegralPart}
U \mkern-20mu \int\limits_{0}^{S^{x=0}(y,z)} \mkern-22mu \deriv x \; \psi^*_i\lef x \rig \psi_f\lef x \rig \equiv \textnormal{SI}_{i,f}\left[ S^{x=0} (y,z) \right],
\end{align}
which is a functional of the SR function. The integral is approximated in Eq.~\ref{matrixElementInfiniteWell} by the width of the integration region multiplied by the value of the integrand at $x=0$. It is a good approximation as long as the integrand is not changing much over the integration region, namely if the SR deviation size is small enough. However, in the limit $U\rightarrow +\infty$ the lowest order gives a finite value while the surface integral is divergent. We can see this more clearly by expanding Eq.~\ref{surfaceIntegralPart} up to second order of the SR function:
\begin{align*}
\textnormal{SI}_{i,f}^{x=0} &\approx  U S^{x=0}(y,z) \left. \psi^*_i (x) \psi_f (x)\right|_{x=0} + U \left[ S^{x=0}(y,z) \right]^2 \frac{\deriv}{\deriv x} \left. \left[ \psi^*_i (x) \psi_f (x) \right] \right|_{x=0} .
\end{align*}
The same result is obtained by rewriting $H^{x=0}$ with $H_0$ in Eq.~\ref{matrixElement}, expanding it up to second order of the SR function and using partial integration to tackle the derivative acting on the Dirac delta function. If we now take the limit $U\rightarrow +\infty$, the first term reaches a finite limit but the second term diverges. It reflects the divergence of an integral over a finite region with an integrand that is infinitely large. The divergence was put under the rug by approximating the wave function by its value at the boundary, where the wave function should become zero when the potential is taken to be infinite.
This analysis shows that approximating the surface integral up to first order of the SR function is not fully compatible with the infinite barrier limit. Therefore the validity of simulation results based on Prange-Nee can rightfully be questioned in some cases. Expanding SI${}_{i,f}$ up to first order of the SR function without an infinite barrier limit, thus requiring an extra parameter $U$, might be a considerable improvement and is considered in section \ref{sectionResultsDiscussion}, referred to as the first order approximation. One could also expand Eq.~\ref{surfaceIntegralPart} up to higher orders of the SR function to obtain a even better approximation. For realistic sizes of SR however, we expect this approach to be impractical because the wave functions can be highly oscillating (see Fig.~\ref{WFInt}) and we would have to go to very high orders to capture the oscillations of the integrand.

Instead of considering higher order expansions of the surface integral SI${}_{i,f}$, the integral in Eq.~\ref{surfaceIntegralPart} can be calculated analytically, taking into account the contribution of every order of the SR function. There are two different solutions, depending on the sign of the SR function, one being the solution for the product of exponential tails, the other for the product of oscillating wave functions inside the rectangular cross section of the wire. Below is an example for the surface integral at the $x=0$ surface with initial state $\mid i \rangle$ and final state $\mid f \rangle$. We write the wave functions (only considering even wave functions with respect to $x=L_x/2$ in the text, odd wave functions can be treated analogously) as:
\begin{align*}
\psi_i (x) &=
\left\{
\begin{matrix}
\mkern-10mu A_i e^{\alpha_i x}, \quad \qquad \qquad \qquad \mkern3mu \textnormal{for } x \leq 0 \qquad \\
B_i \cos \left[ k_i \lef x - L_x/2 \rig \right], \; \; \mkern2mu \textnormal{for } 0\leq x \leq L_x \\
\mkern3mu A_i e^{-\alpha_i (x - L_x)}, \quad \quad \qquad \; \; \textnormal{for } x \geq L_x \qquad
\end{matrix} \right. .
\end{align*}
The exponential tail on the other side of the wire in the integral of Eq.~\ref{surfaceIntegralPart} can safely be neglected because we assumed that the SR function is never larger than the wire width or height. The real constants $A_i$, $B_i$, $\alpha_i$ and $k_i$ are obtained from the finite potential well solution in the $x$-direction.
Insertion into Eq.~\ref{surfaceIntegralPart} leads to:
\begin{widetext}
\begin{align}
\label{surfaceIntFunctions}
\textnormal{SI}_{i,f}\left[S^{x=0}\right] &\equiv \left\{ \begin{matrix}
                                       \textnormal{SI}^{+}_{i,f}\left[S^{x=0}\right], \quad S^{x=0} > 0 \\
				       \textnormal{SI}^{-}_{i,f}\left[S^{x=0}\right], \quad S^{x=0} < 0
                                      \end{matrix} \right. ,
\quad \textnormal{SI}^-_{i,f}\left[S^{x=0}\right] \equiv -U A_i A_f \frac{1 - e^{(\alpha_i + \alpha_f) S^{x=0}}}{\alpha_i + \alpha_f} \\ \notag
\textnormal{SI}^+_{i,f}\left[S^{x=0}\right] & \equiv U\frac{B_i B_f}{2} \sum\limits_{\pm} \frac{\sin \left[\lef k_f \pm k_i \rig L_x / 2 \right] - \sin \left[ \lef k_f \pm k_i \rig \lef L_x - 2 S^{x=0} \rig / 2 \right]}{k_f \pm k_i}.
\end{align}
\end{widetext}
Next, these solutions have to be inserted into the SR matrix element given by Eq.~\ref{exactMatrixElement} and the average has to be taken over the absolute value squared of the matrix element appearing in the BTE (Eq.~\ref{RTA}). We obtain for the contribution of the surface near $x=0$:
\begin{align}
\label{avSquareMatrixEl}
& \left\langle \left| \left\langle i \mid V^{x=0} \mid f \right\rangle \right|^2 \right\rangle \\ \notag
&\; = \frac{1}{L_z^2} \int\limits_{0}^{L_y} \mkern-3mu \deriv y \mkern-8mu \int\limits_{-L_z/2}^{+L_z/2} \mkern-12mu \deriv z \int\limits_{0}^{L_y} \mkern-3mu \deriv \tilde{y} \mkern-8mu \int\limits_{-L_z/2}^{+L_z/2} \mkern-12mu \deriv \tilde{z} \; \left\langle \textnormal{SI}_{i,f}^{x=0} \tilde{\textnormal{SI}}_{i,f}^{x=0} \right\rangle \psi_i (y) \psi_f (y) \psi_f (\tilde{y}) \psi_i (\tilde{y}) e^{-i (k^z_i - k^z_f) (z - \tilde{z})}.
\end{align}
Although the SR of different boundary surfaces is considered to be uncorrelated, we do observe couplings between them in the average of the absolute squared of the matrix element. This is because $\textnormal{SI}\left[ S \right]$ is in general not symmetric around $S=0$ thus leaving the average different from zero. Below the general form of the coupling between the $x=0$ and $y=0$ surface is given:
\begin{align}
\label{avSquareMatrixElAdiag}
& \left\langle \langle i \mid V^{x=0} \mid f \rangle \langle f \mid V^{y=0} \mid i \rangle \right\rangle \\ \notag
&\; = \frac{1}{L_z^2} \int\limits_{0}^{L_y} \mkern-3mu \deriv y \mkern-8mu \int\limits_{-L_z/2}^{+L_z/2} \mkern-12mu \deriv z \int\limits_{0}^{L_x} \mkern-3mu \deriv \tilde{x} \mkern-8mu \int\limits_{-L_z/2}^{+L_z/2} \mkern-12mu \deriv \tilde{z} \; \left< \textnormal{SI}_{i,f}^{x=0} \tilde{\textnormal{SI}}_{i, f}^{y=0} \right> \psi_i (y) \psi_f (y) \psi_f (\tilde{x}) \psi_i (\tilde{x}) e^{-i (k^z_i - k^z_f) (z - \tilde{z})}.
\end{align}

Inserting the expressions from Eq.~\ref{surfaceIntFunctions} in Eq.~\ref{avSquareMatrixEl} or Eq.~\ref{avSquareMatrixElAdiag}, the average over the SR functions can be split up in different regions depending on their sign:
\begin{align}
\label{averageSurface}
\left\langle \textnormal{SI}_{i,f} \tilde{\textnormal{SI}}_{i,f} \right\rangle &= \left\langle \textnormal{SI}^{+}_{i,f} \tilde{\textnormal{SI}}^{+}_{i,f} \right\rangle + 2 \Re \lef \left\langle \textnormal{SI}^{+}_{i,f} \tilde{\textnormal{SI}}^{-}_{i,f} \right\rangle \rig + \left\langle \textnormal{SI}^{-}_{i,f} \tilde{\textnormal{SI}}^{-}_{i,f} \right\rangle, \\ \notag
\left\langle \textnormal{SI}_{i,f} \right\rangle &= \left\langle \textnormal{SI}^+_{i,f} \right\rangle + \left\langle \textnormal{SI}^-_{i,f} \right\rangle.
\end{align}
To continue, the average over the SR function in the surface integral should be calculated. This requires a distribution function $f (S, \tilde{S})$ that is consistent with the Gaussian autocorrelation function of Eq.~\ref{gaussianCorr}. There are several ways to obtain an average that is consistent with the Gaussian autocorrelation and three approaches are discussed in the following subsections.

\subsubsection{Bivariate normal model}
\begin{figure}[tb]\begin{center}
\subfigure[\ $S>0$, $\tilde{S}>0$]{
\includegraphics[width=0.35\linewidth]{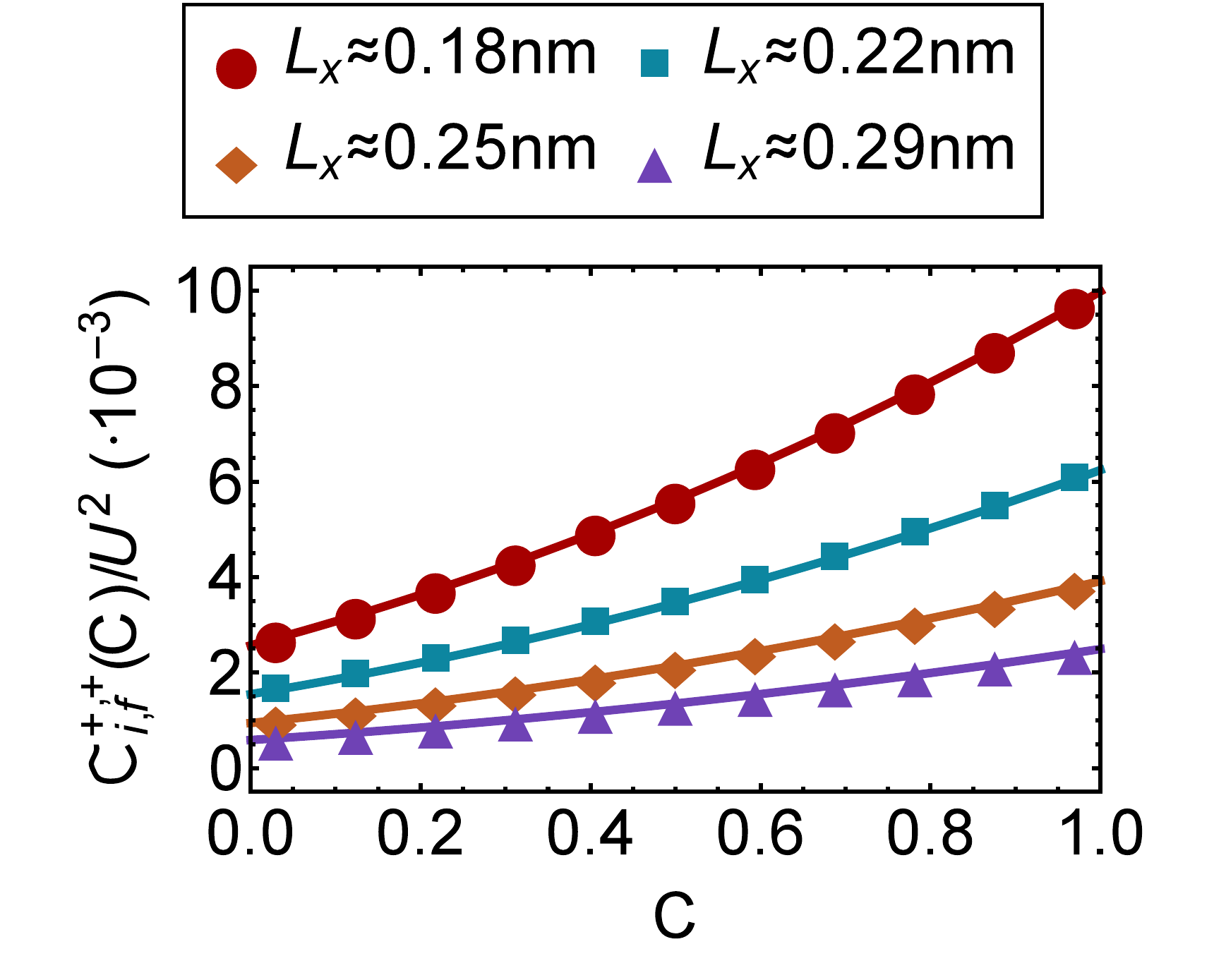}}
\subfigure[\ $S>0$, $\tilde{S}<0$]{
\includegraphics[width=0.35\linewidth]{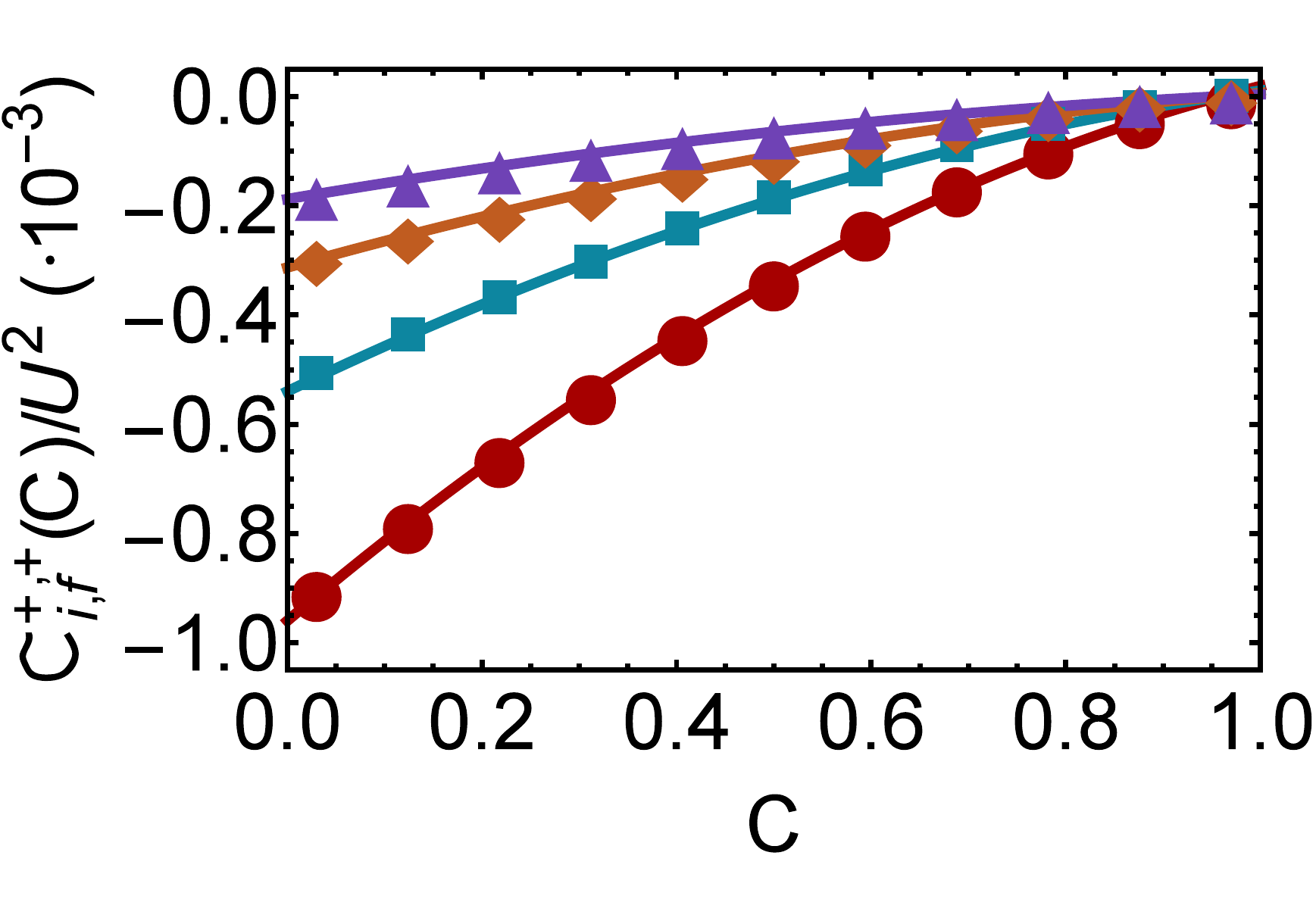}}
\subfigure[\ $S<0$, $\tilde{S}>0$]{
\includegraphics[width=0.35\linewidth]{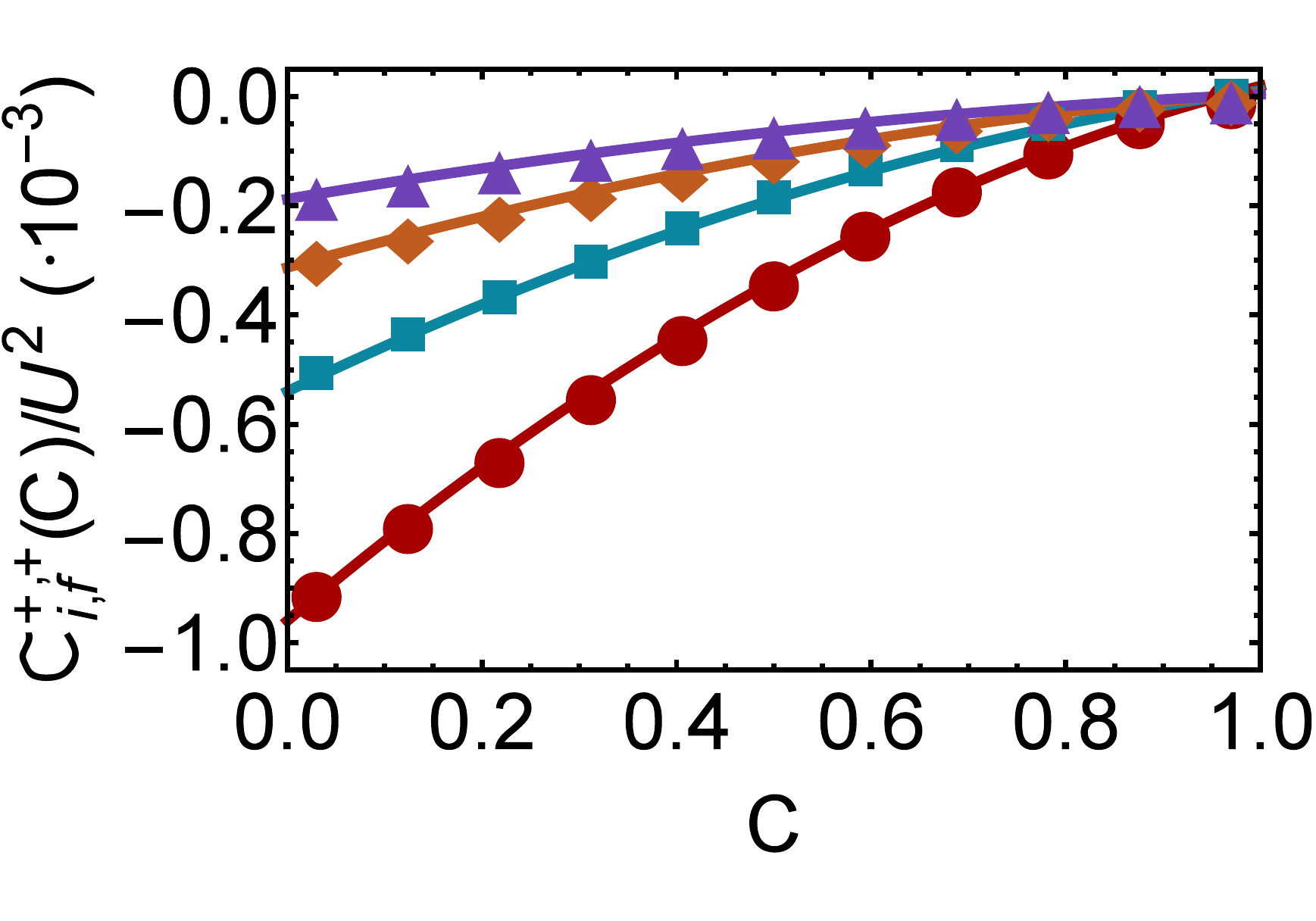}}
\subfigure[\ $S<0$, $\tilde{S}<0$]{
\includegraphics[width=0.35\linewidth]{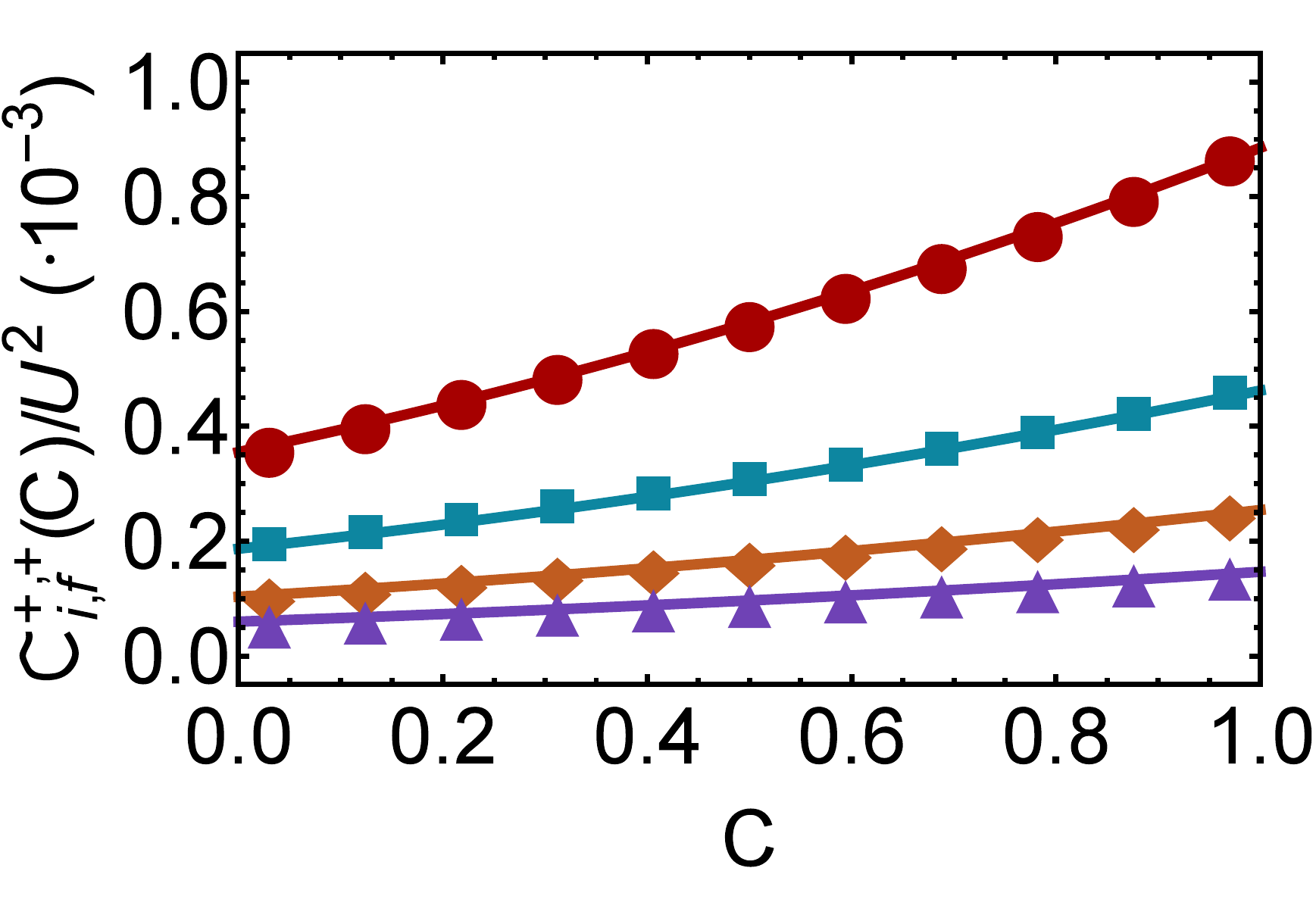}}
\end{center}
\caption{Using Eq.~\ref{bivNormalFit} to fit the function $\mathcal{C}_{i,f}(C)$ in the different quadrants of the $(S,\tilde{S})$-plane for wire simulations with $L_x$ ranging from 0.18 to 0.29~nm and $\Delta \approx 0.036$~nm, $\Lambda \approx 0.11$~nm. These nanowire dimensions are unrealistic and relate to the single subband toy model discussed in section~\ref{SakakiModel}.}
\label{fitBivariate}
\end{figure}

We can suppose a bivariate normal distribution function, as was done recently by Lizzit et al.\cite{lizzit2014new} For the $x=0$ surface this gives:
\begin{align}
\label{bivariateNormal}
\left\langle \textnormal{SI}_{i,f}^{x=0} \tilde{\textnormal{SI}}_{i,f}^{x=0} \right\rangle &= \int\limits_{-\infty}^{+\infty} \mkern-5mu \deriv S \int\limits_{-\infty}^{+\infty} \mkern-5mu \deriv \tilde{S} \; \frac{\textnormal{SI}_{i, f}^{x=0}\left[ S \right] \textnormal{SI}_{i, f}^{x=0} [\tilde{S}]}{2\pi \Delta^2\sqrt{1-C\lef \mathbf{r}, \tilde{\mathbf{r}} \rig^2}} \exp\left[ -\frac{S^2 + \tilde{S}^2  - 2 C\lef \mathbf{r}, \tilde{\mathbf{r}} \rig S \tilde{S}}{2 \Delta^2 \left[ 1 - C\lef \mathbf{r}, \tilde{\mathbf{r}} \rig^2 \right]} \right], \\ \notag
C\lef \mathbf{r}, \tilde{\mathbf{r}} \rig &\equiv e^{-[(y - \tilde{y})^2 + (z - \tilde{z})^2 / (\Lambda^2/2)}.
\end{align}
As the latter integral cannot be evaluated analytically, it needs to be computed numerically for every pair $i,f$ and every possible value of $0\leq C(\mathbf{r},\tilde{\mathbf{r}}) \leq 1$. The integration of Eq.~\ref{bivariateNormal} over the quadrant in the $(S,\tilde{S})$-plane with $S,\tilde{S}>0$, denoted by the functional $\mathcal{C}_{i,f}^{+,+}\left[ C(\mathbf{r},\tilde{\mathbf{r}}) \right]$, can be fitted as a power series of $C(\mathbf{r}, \tilde{\mathbf{r}})$:
\begin{align}
\label{bivNormalFit}
& \mathcal{C}^{+,+}_{i,f} \left[ C \lef \mathbf{r}, \tilde{\mathbf{r}} \rig \right] \approx c_0^{+,+} + c_1^{+,+} C\lef \mathbf{r}, \tilde{\mathbf{r}} \rig + c_2^{+,+} C^2\lef \mathbf{r}, \tilde{\mathbf{r}} \rig + \ldots,
\end{align}
with fitting parameters $c_{0,1,2,\ldots}^{+,+}$. Fitting parameters $c_{0,1,2,\ldots}^{\pm,\tilde{\pm}}$ can be found for the integration over every quadrant of the $(S,\tilde{S})$-plane in the same way, with the superscripts $\pm$ and $\tilde{\pm}$ representing the sign of $S$ and $\tilde{S}$ respectively. The corresponding fitting function can be plugged in the remaining integrals of e.g. Eq.~\ref{avSquareMatrixEl}, such that the remaining integration over the boundary coordinates can be done analytically. The fitting function up to second order of $C$ appears to capture the behavior quite well (see Fig.~\ref{fitBivariate}).

For some of the pairs $i,f$, the integrand in Eq.~\ref{bivariateNormal} is highly oscillating and the number of pairs increases rapidly for larger diameters, so it proves to be a tedious numerical job. We have not been able to optimize this method for metallic wire simulations with width or height up to 10~nm (see Fig.~\ref{compTime} for a comparison of computation times). To overcome this problem of numerical integration, we propose two other distribution functions that are consistent with the Gaussian autocorrelation (for $S$ and $\tilde{S}$ SR functions of the same boundary) below.

\subsubsection{Finite domain model I}
\label{sectionFin1}
\begin{figure}[tb]\centering
\includegraphics[width=0.4\linewidth]{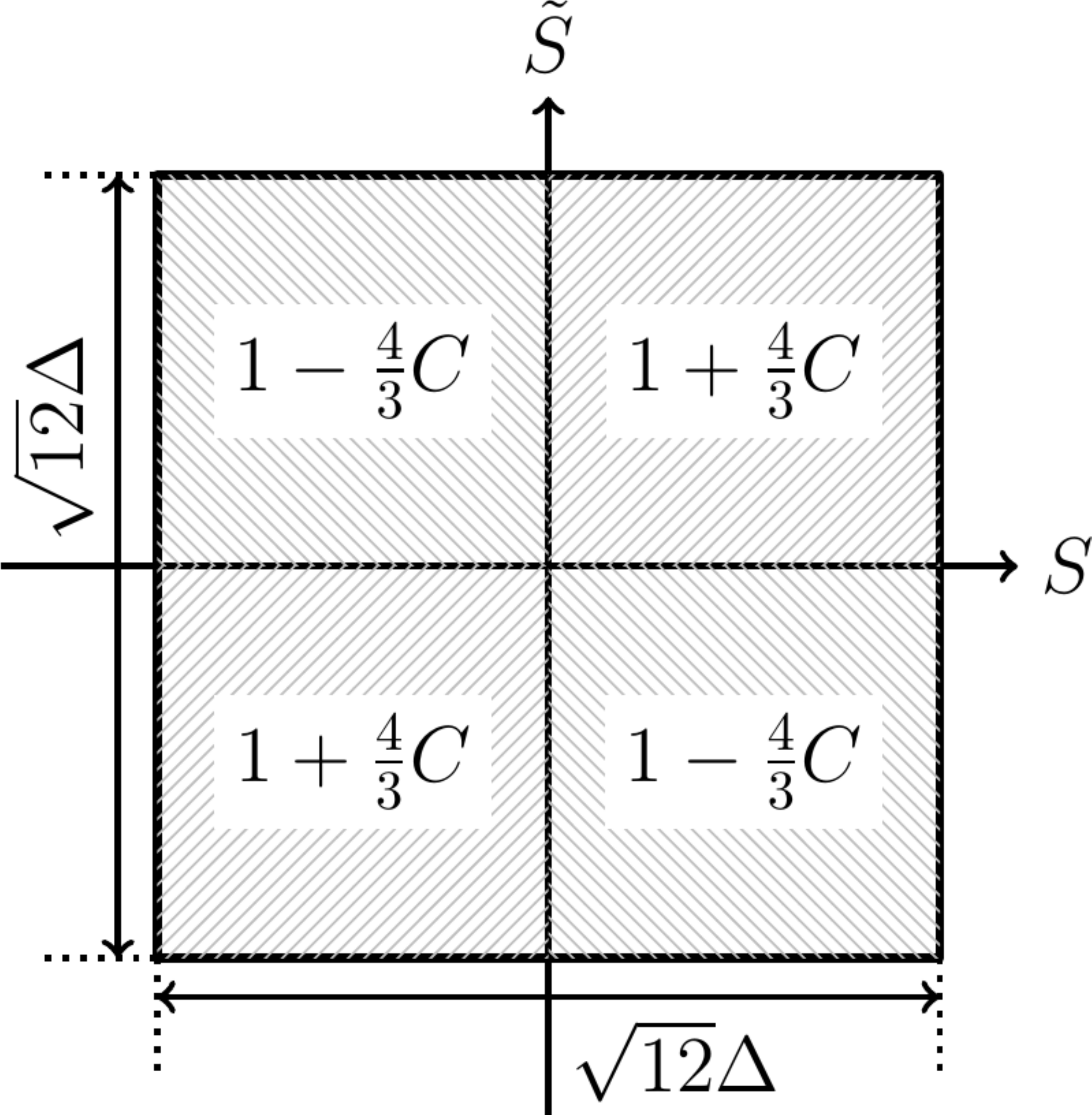}
\caption{The finite domain for $S$ and $\tilde{S}$ is $\left[ -\sqrt{3}\Delta, +\sqrt{3}\Delta \right] \times \left[ -\sqrt{3}\Delta, +\sqrt{3}\Delta \right]$ and the weight of $f (S, \tilde{S})$ is given by $1+4C\lef \mathbf{r}, \tilde{\mathbf{r}} \rig/3$ and $1-4C\lef \mathbf{r}, \tilde{\mathbf{r}} \rig/3$ in the equal sign and unequal sign quadrants respectively. It is not difficult to check that this is consistent with the Gaussian autocorrelation function $C\lef \mathbf{r}, \tilde{\mathbf{r}} \rig$. For $S$ and $\tilde{S}$ SR functions corresponding to different boundaries, $C\lef \mathbf{r}, \tilde{\mathbf{r}} \rig$ is considered to be zero (no correlation).}
\label{finiteDomain}
\end{figure}

An approach that allows for an analytical expression of the average over SR functions consists of considering uniformly distributed variables over a finite interval around the average position of the wire boundary surface. The finite interval also makes sure that no outliers of the SR function are passing the other side of the wire, consistent with the assumption in Eq.~\ref{surfaceIntFunctions} that a rough boundary does not cross the other side of the wire, unlike the bivariate normal distribution function in Eq.~\ref{bivariateNormal}.
A standard deviation and correlation length is introduced in the distribution function $f (S, \tilde{S})$ by changing the domain size and assigning different weights to different quadrants of the $(S,\tilde{S})$ plane respectively, as shown in Fig.~\ref{finiteDomain}, while retaining the appropriate mean value of the SR functions ($\langle S \rangle=0$).
\setlength{\unitlength}{1.6cm}
Using this model for the average in Eq.~\ref{averageSurface}, we get:
\begin{align}
\label{averageSurface1}
& \left\langle \textnormal{SI}_{i,f} \tilde{\textnormal{SI}}_{i,f} \right\rangle = \left[ 1 + 4C\lef \mathbf{r}, \tilde{\mathbf{r}} \rig/3 \right] \left\langle \textnormal{SI}_{i,f} \tilde{\textnormal{SI}}_{i,f} \right\rangle^{\textnormal{diag.}} + \left[ 1 - 4C\lef \mathbf{r}, \tilde{\mathbf{r}} \rig/3 \right] \left\langle \textnormal{SI}_{i,f} \tilde{\textnormal{SI}}_{i,f} \right\rangle^{\textnormal{off-diag.}}, \\ \notag
&\left\langle \textnormal{SI}_{i,f} \tilde{\textnormal{SI}}_{i,f} \right\rangle^{\textnormal{diag.}} \equiv \frac{1}{12\Delta^2} \lef \left| \int\limits_0^{\sqrt{3}\Delta} \mkern-10mu \deriv S \; \textnormal{SI}^{+}_{i,f}\left[ S \right] \right|^2 + \left| \mkern5mu \int\limits^0_{-\sqrt{3}\Delta} \mkern-12mu \deriv S \; \textnormal{SI}^{-}_{i,f}\left[ S \right] \right|^2 \rig,
\end{align}
\begin{align*}
&\left\langle \textnormal{SI}_{i,f} \tilde{\textnormal{SI}}_{i,f} \right\rangle^{\textnormal{off-diag.}} \equiv \frac{1}{12\Delta^2} \lef \int\limits_0^{\sqrt{3}\Delta} \mkern-10mu \deriv S \; \textnormal{SI}^{+}_{i,f}\left[ S \right] \mkern-10mu \int\limits^0_{-\sqrt{3}\Delta} \mkern-12mu \deriv \tilde{S} \; \textnormal{SI}^{-}_{i,f} \left[ \tilde{S} \right] + \mkern-10mu \int\limits^0_{-\sqrt{3}\Delta} \mkern-12mu \deriv S \; \textnormal{SI}^{-}_{i,f}\left[ S \right] \mkern-3mu \int\limits_0^{\sqrt{3}\Delta} \mkern-10mu \deriv \tilde{S} \; \textnormal{SI}^{+}_{i,f}\left[ \tilde{S} \right] \rig.
\end{align*}
When looking at Eq.~\ref{surfaceIntFunctions}, it is clear that only an integration over exponentials with complex argument remains, while applying the bivariate normal distribution (Eq.~\ref{bivariateNormal}) leads to a remaining integral over a product of exponential, Gaussian and error functions for which an analytical expression could not be found.

A downside of this approach is the unphysical negative weight that is attributed to the anti-correlated quadrants when $C(\mathbf{r},\tilde{\mathbf{r}})>3/4$. This should be kept in mind when using this approach to simplify the integrals over the SR functions.

\subsubsection{Finite domain model II}
\label{sectionFin2}
\begin{figure}[tb]
\begin{center}
\includegraphics[width=0.4\linewidth]{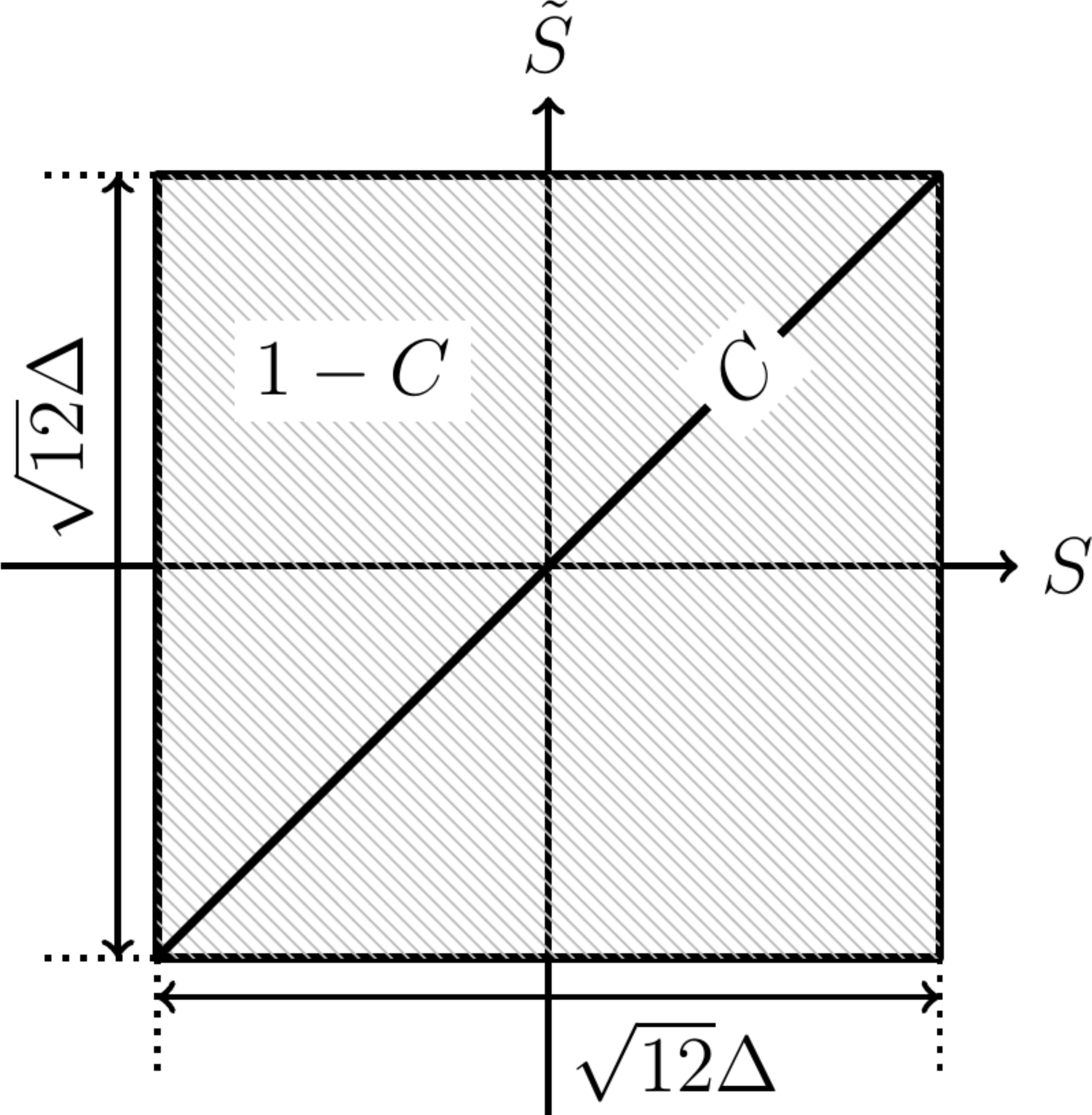}
\end{center}
\caption{The finite domain for $S$ and $\tilde{S}$ is $\left[ -\sqrt{3}\Delta, +\sqrt{3}\Delta \right] \times \left[ -\sqrt{3}\Delta, +\sqrt{3}\Delta \right]$ and the weight of this domain is $1-C\lef \mathbf{r}, \tilde{\mathbf{r}} \rig$. The correlated part is the diagonal $S=\tilde{S}$ with $-\sqrt{3}\Delta < S < \sqrt{3}\Delta$ and weight $C\lef \mathbf{r}, \tilde{\mathbf{r}} \rig$. This model is also consistent with the Gaussian autocorrelation function. For $S$ and $\tilde{S}$ SR functions corresponding to different boundaries, $C\lef \mathbf{r}, \tilde{\mathbf{r}} \rig$ is considered to be zero.}
\label{finiteDomain2}
\end{figure}

Instead of assigning different weights to different quadrants, the model proposed in this section consists of a sum of a fully correlated part and a fully uncorrelated part. This also makes the average over the SR functions analytically solvable:
\begin{align}
\label{averageSurface2}
& \left\langle \textnormal{SI}_{i,f} \tilde{\textnormal{SI}}_{i,f} \right\rangle = C\lef \mathbf{r}, \tilde{\mathbf{r}} \rig \left\langle \textnormal{SI}_{i,f} \textnormal{SI}_{i,f} \right\rangle^{\textnormal{corr.}} + \left[ 1 - C\lef \mathbf{r}, \tilde{\mathbf{r}} \rig \right] \left\langle \textnormal{SI}_{i,f} \tilde{\textnormal{SI}}_{i,f} \right\rangle^{\textnormal{uncorr.}}, \\ \notag
&\left\langle \textnormal{SI}_{i,f} \tilde{\textnormal{SI}}_{i,f} \right\rangle^{\textnormal{corr.}} \equiv \frac{1}{\sqrt{12}\Delta} \int\limits_{-\sqrt{3}\Delta}^{\sqrt{3}\Delta} \mkern-12mu \deriv S \; \left| \textnormal{SI}_{i,f}\left[ S \right] \right|^2, \quad
\left\langle \textnormal{SI}_{i,f} \tilde{\textnormal{SI}}_{i,f} \right\rangle^{\textnormal{uncorr.}} \equiv \frac{1}{12\Delta^2} \left| \mkern3mu \int\limits_{-\sqrt{3}\Delta}^{\sqrt{3}\Delta} \mkern-12mu \deriv S \; \textnormal{SI}_{i,f}\left[ S \right] \right|^2.
\end{align}
The uncorrelated part has a uniform distribution in the $(S,\tilde{S})$-plane, whereas the correlated part has a distribution function that only allows identical values of $S$ and $\tilde{S}$, as shown in Fig.~\ref{finiteDomain2}. The full SR matrix element expression of this model and finite domain model I can be found in appendix \ref{sectionAppendix}.
\setlength{\unitlength}{1.6cm}

\begin{figure}[tb]\begin{center}
\subfigure[\ $0 \leq C \leq 1$]{
\includegraphics[width=0.35\linewidth]{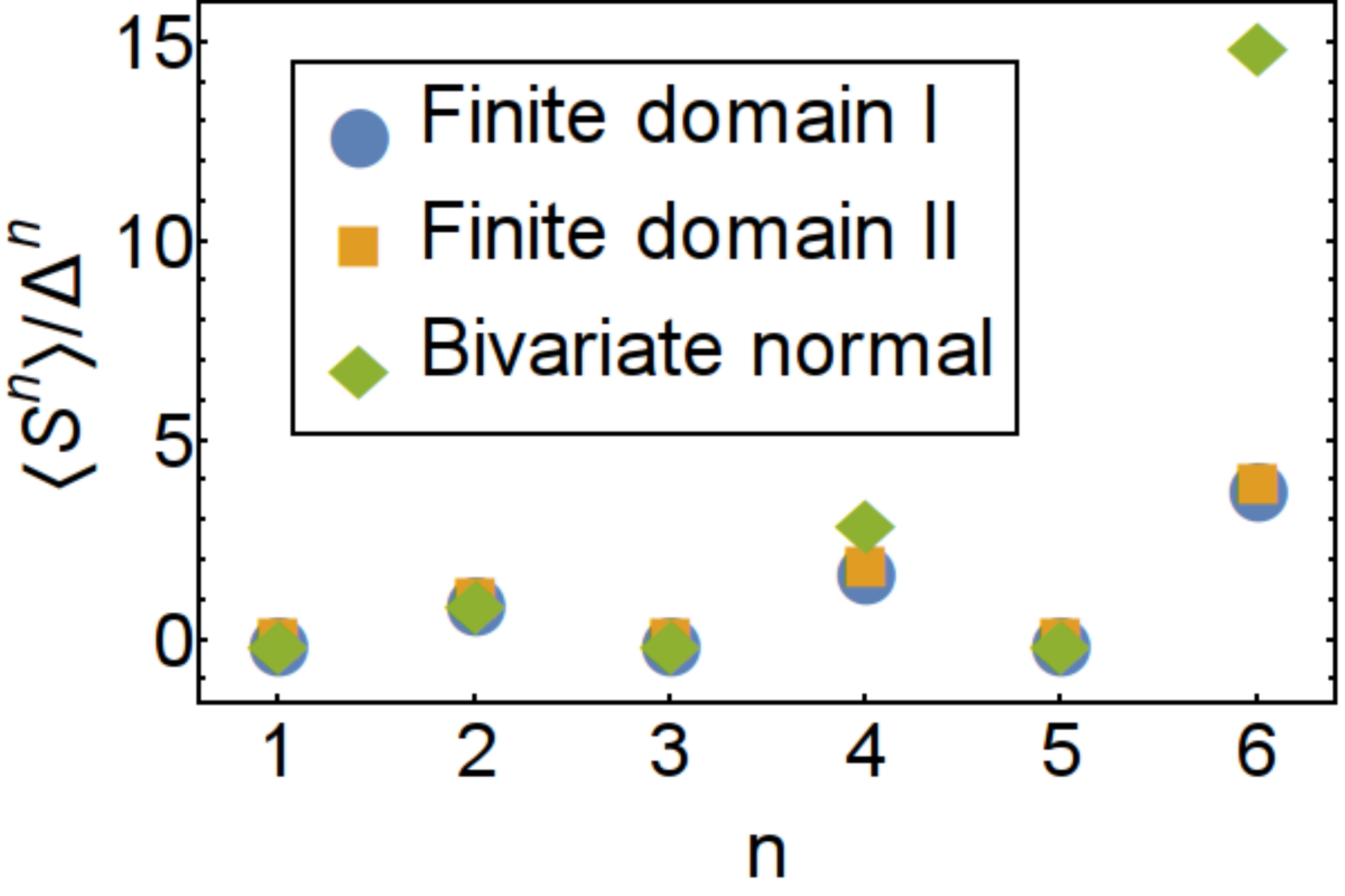}}
\subfigure[\ $C = 1$]{
\includegraphics[width=0.35\linewidth]{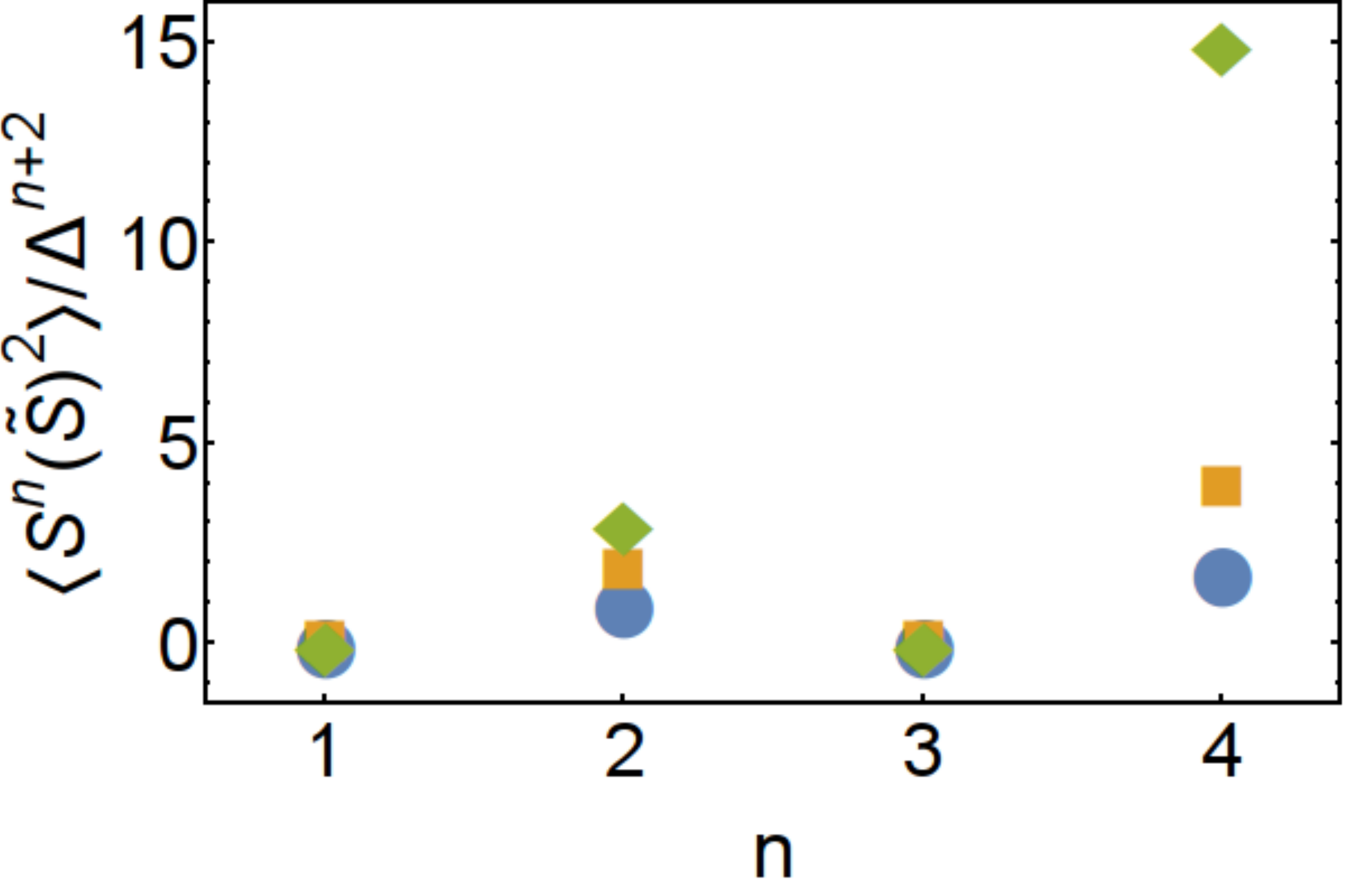}}
\subfigure[\ $C = 0.25$]{
\includegraphics[width=0.35\linewidth]{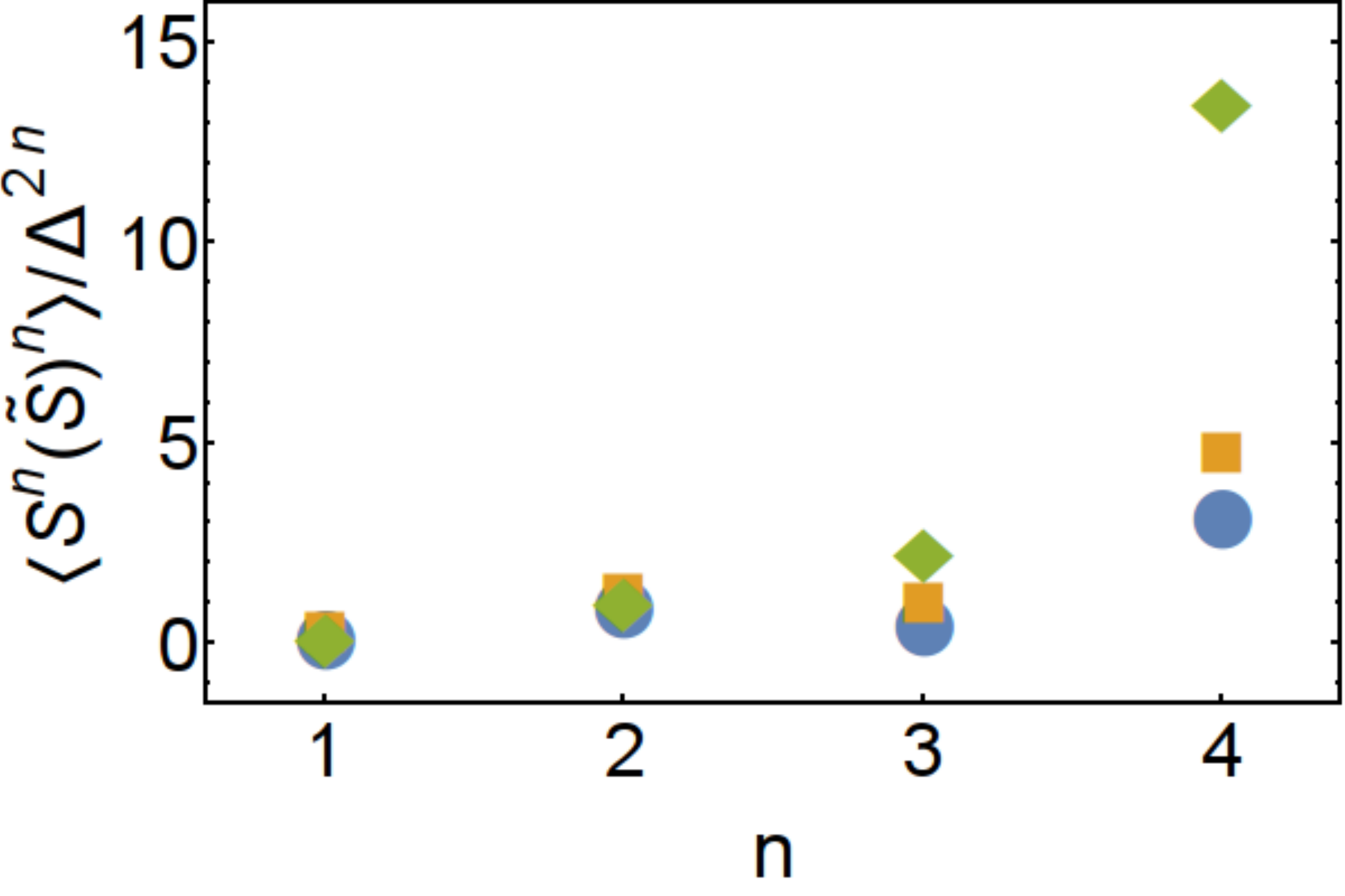}}
\subfigure[\ $C = 0.75$]{
\includegraphics[width=0.35\linewidth]{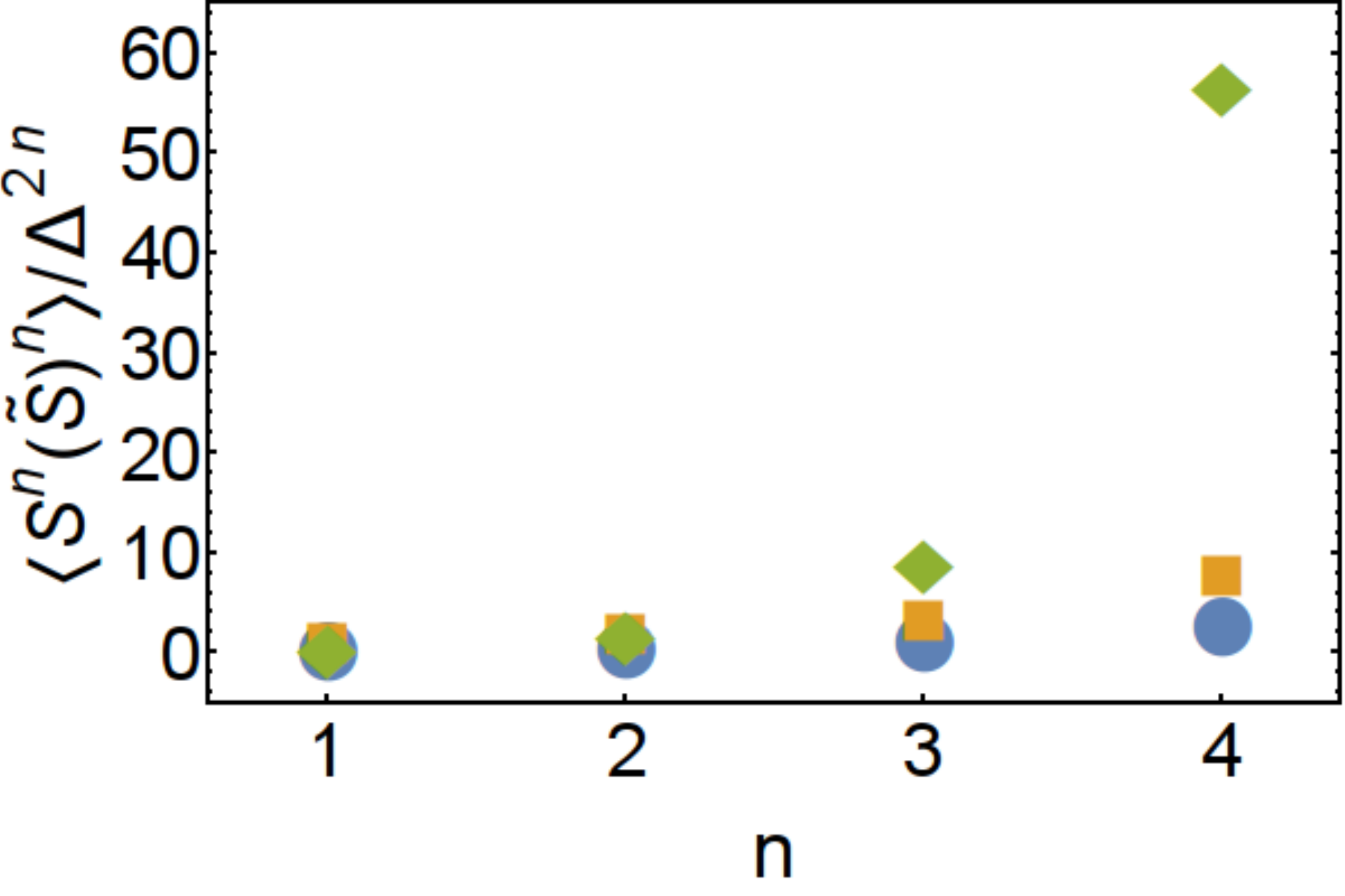}}
\end{center}
\caption{The expectation values of products of SR functions are shown for several values of correlation $C$, using the different distribution functions that are proposed in section \ref{correctSolutionSection}.}
\label{modelComparisonModes}
\end{figure}

Even though this model and the finite domain model I are consistent with the bivariate normal distribution function for $\langle S^m \tilde{S}^{n} \rangle$ with $m+n \leq 2$, the higher moments ($m+n>2$) will be different. The higher moments of this model and the other models are compared in Fig.~\ref{modelComparisonModes}. They are different for each distribution function and the differences become more substantial when higher values of $n$ or $m$ are considered. The finite domain models agree quite well but the expectation values deviate substantially from the bivariate normal model. The expectation values are in general larger for the latter because higher values of the SR function contribute (infinite domain). All the distribution functions respect the symmetry and give zero when $n+m$ is odd.

\section{Results \& Discussion}
\label{sectionResultsDiscussion}

\subsection{Single subband toy model}
\label{SakakiModel}
\begin{figure}[tb]\begin{center}
\subfigure [\ Single subband toy model] {
\includegraphics[width=0.4\linewidth]{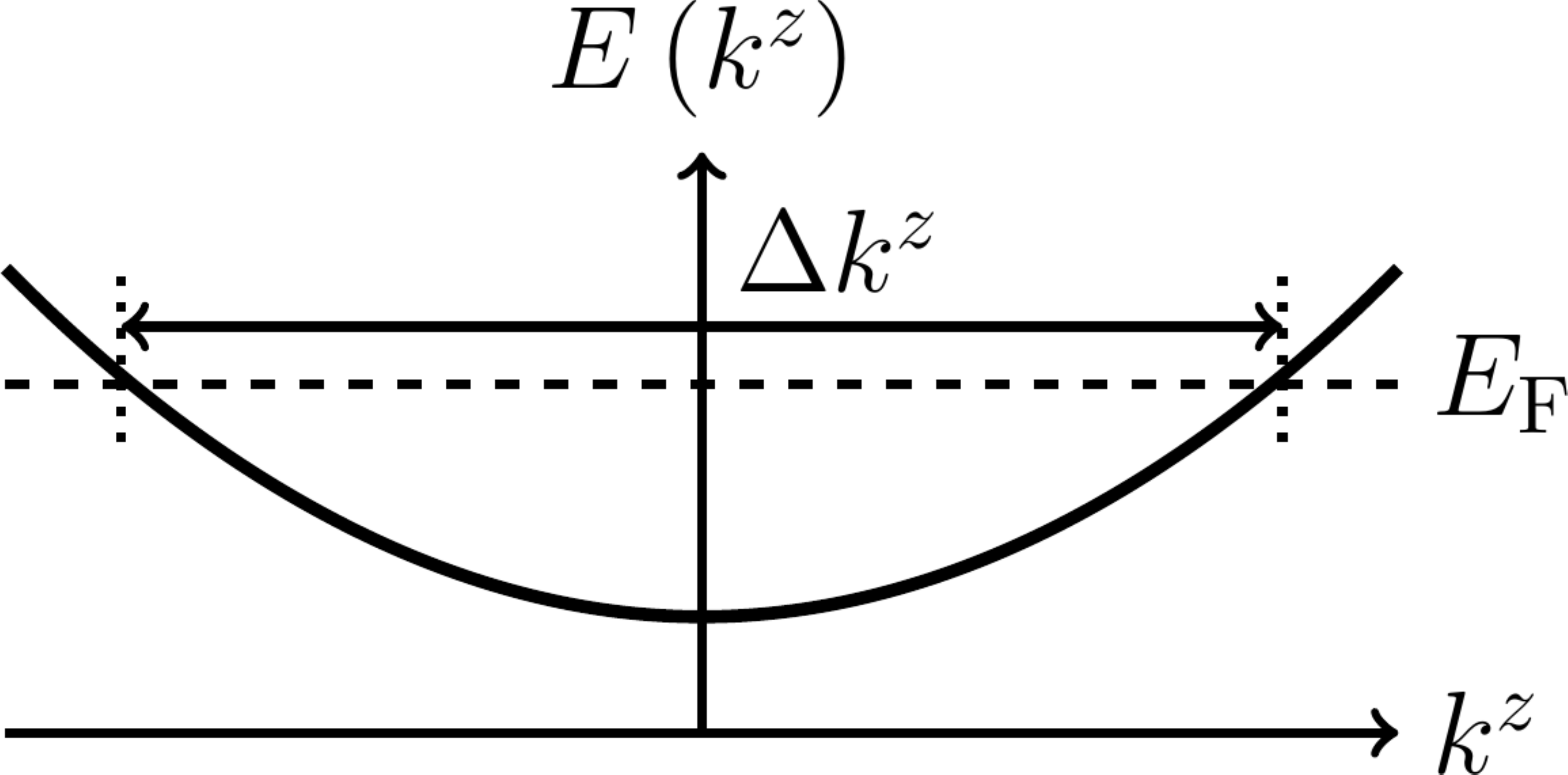}
}
\subfigure[\ Resistivity]{
\includegraphics[width=0.4\linewidth]{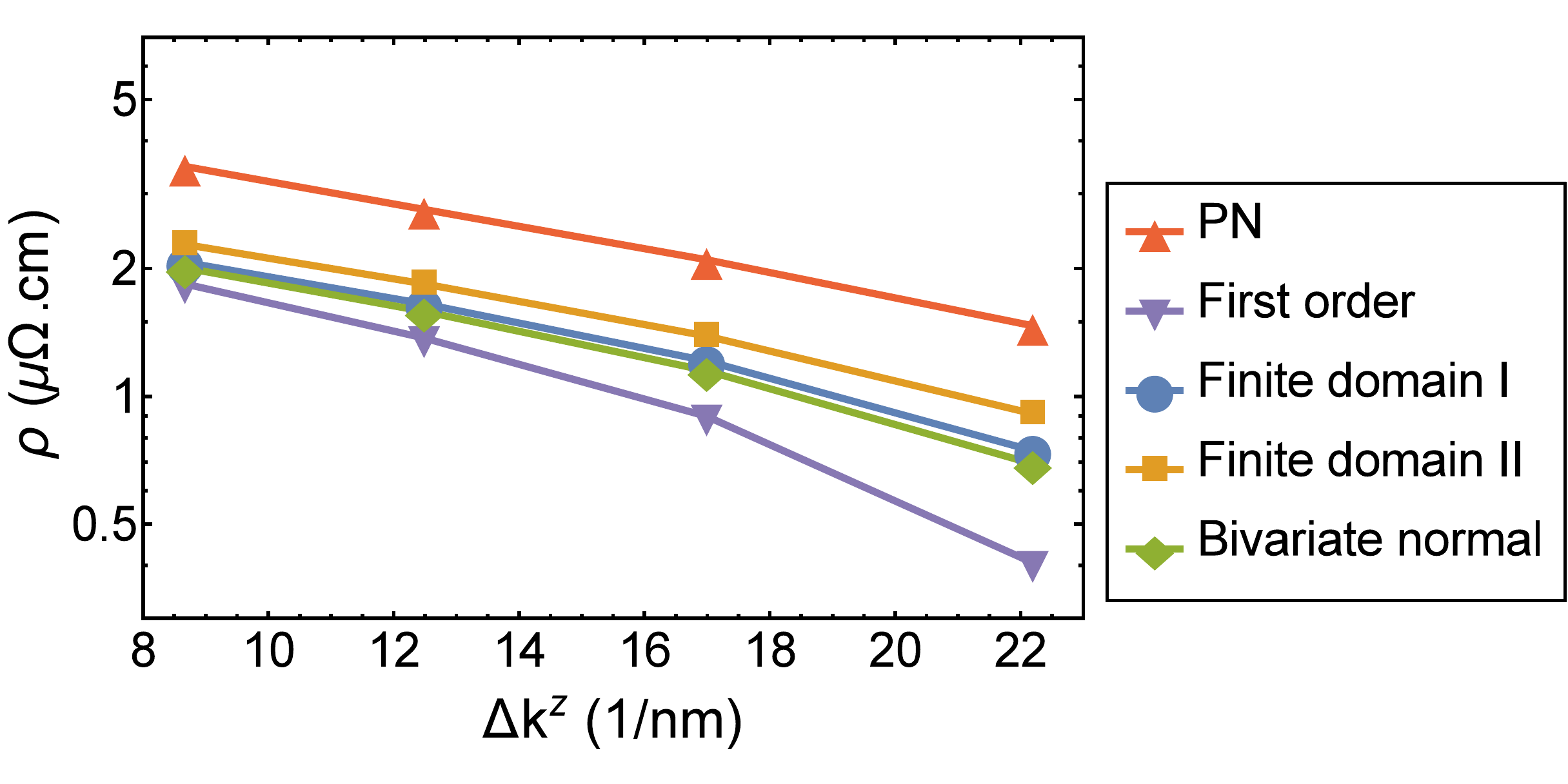}}
\end{center}
\caption{In (a) the dispersion relation of a single subband toy model is shown with the momentum gap $\Delta k^z$ indicated. The resistivity values obtained with SR matrix elements are shown in (b) in a log scale, using the Prange-Nee model (red triangles), first order model (purple upside-down triangles), finite domain models (I: blue circles, II: yellow squares) and finally the bivariate normal model (green diamonds). The simulated wire has a square cross section with width and height ranging from approximately 0.25~nm to 0.29~nm and roughness properties are: $\Delta \approx 0.036$~nm, $\Lambda \approx 0.11$~nm. The Fermi level is obtained by fixing the density to the bulk conduction electron density in copper (7~eV in bulk), but together with the very low width and height, fixing the Fermi level in this way is just a tool to obtain a single subband model and does not represent a realistic nanowire.}
\label{sakakiFig}
\end{figure}

A test case model for the finite domain models proposed in previous section is a wire with only a single subband crossing the Fermi level. This can be realized with a low effective mass, conduction electron density or diameter. While in principle being possible (and already being proposed and investigated\cite{sakaki1980scattering,sakaki1986physical,motohisa1992interface}) for semiconductor nanowires, it is practically impossible to realize for metal nanowires. It serves as a good test case however because there are only two states at the Fermi level hence scattering is limited to Umklapp scattering and we can clearly see the differences of the different methods and the effect of the momentum gap $\Delta k^z$ on the transport. We simulate the single subband model by considering a very narrow wire and by fixing the Fermi level with the bulk electron density. In Fig.~\ref{sakakiFig} the resistivity is shown, using all the methods introduced in section \ref{sectionSRS}. Fig.~\ref{sakakiFig} clearly shows the large overestimation of the Prange-Nee model, more or less by an order of magnitude. The results are very different because we are simulating an extremely narrow nanowire. The infinite barrier approximation is not working well because there is substantial wave function penetration into the potential barrier region. The other methods are much better in agreement, with very good agreement between the bivariate normal model and finite domain model I. The first order approximation of the wave function surface integral does also not differ much different from the former two models. The first order approximation is expected to work well for this toy model because there is only one subband and therefore the wave function, having a large wave vector, varies little over the SR standard deviation. A first order approximation of the integral in Eq.~\ref{exactMatrixElement} is acceptable in this case. The approximation is expected to be worse when more subbands (and higher wave vectors) come into play.

The increase of resistivity $\rho$ for decreasing $\Delta k^z$ seems to follow a $\rho = \rho_\textnormal{max.} \exp\lef - L \Delta k^z \rig$ scaling quite well for the Prange-Nee model, but the other methods deviate from this with an even larger suppression for large $\Delta k^z$. The parameter $\rho_\textnormal{max.}$ differs from method to method, but all the methods agree quite well on a parameter $L\approx 0.1$~nm. Although the scaling of Eq.~\ref{critMomentumGapPN} is different from the scaling in the simulation results, both indicate an exponential suppression for large $\Delta k^z$ and it gives a critical length scale $1/\Delta k^z_\textnormal{crit. (PN)} \approx 0.04$~nm of the same order of magnitude as $L$ obtained from the fit. This behavior is very important because every wire is characterized by a minimal $\Delta k^z$ between positive and negative wave vector states and it appears to have a crucial impact on the self-consistent solution of relaxation times when more subbands are considered. Hence, it also crucially affects the resistivity, as will be discussed in more detail in subsection \ref{sectionNanowires}.

\setlength{\unitlength}{1.6cm}
\begin{figure}[tb]\centering
\includegraphics[width=0.35\linewidth]{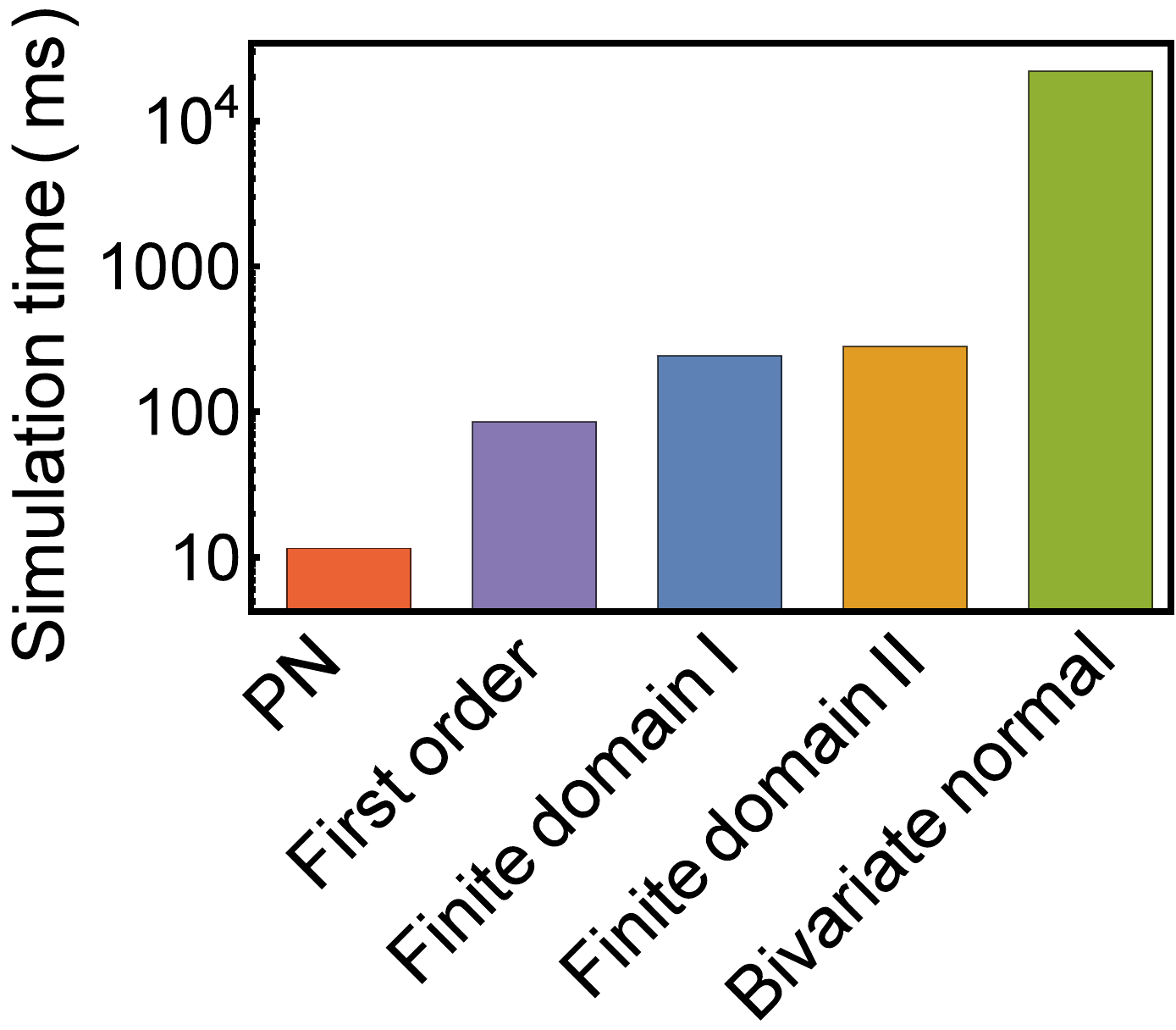}
\caption{The simulation time for the different methods used to calculate the SR matrix elements is shown on a logarithmic scale (in milliseconds) for the single subband simulation of the four data points presented in Fig.~\ref{sakakiFig}~(b). The simulation is performed on a single i5 3.4GHz CPU and is based on a global adaptive numerical simulation strategy. A realistic nanowire size would increase the number of computations by a factor of $10^4$ to $10^5$.}
\label{compTime}
\end{figure}

Other than the resistivity values of the different methods, there is also the difference in computational burden. We expect the Prange-Nee model to be the fastest and the numerical integration of the bivariate normal model to be the slowest. This is confirmed by comparing the simulation times in Fig.~\ref{compTime}. Currently the simulations of the finite domain models are more or less one order of magnitude slower than the Prange-Nee model, while the bivariate normal model performs even slower by two orders of magnitude. While it must be said that our implementation of the numerical integration might be further optimized to enhance the computation speed of the latter, it will be hard to beat the performance of simulations using finite domain model I or II because they only consist of evaluations of analytical expressions.

\subsection{Metallic nanowires}
\label{sectionNanowires}
In order to simulate realistic metallic nanowires, the model parameters have to be optimally determined. For the barrier height we consider the value corresponding to vacuum using the relation $U = E_\textnormal{\tiny F} + W$, with $W$ the work function and $E_\textnormal{\tiny F}$ the Fermi energy for bulk.\cite{kittel1976introduction} The states of the subbands that cross the Fermi level are obtained self-consistently for every wire dimension by filling up all the subbands up to the appropriate carrier density of the considered material. All the simulations rely on the underlying assumption of a close to rectangular cross section and an (isotropic) effective mass for the conduction electrons (taken from measured bulk values), which can rightfully be questioned for very narrow metallic nanowires and some materials. However, the presented approach with finite domain distribution functions and their analytical expressions are still applicable to more detailed wave functions and subband structures (obtained with ab initio methods for example), as long as the wire boundary consists of flat surfaces. In this case, one can define SR functions on those surfaces and apply Eq.~\ref{averageSurface1}-\ref{averageSurface2} straightforwardly by expanding the wave functions on a basis of exponentials or sines and cosines.\footnote{For cylindrical nanowires, it is better to start from circular potential well solutions and to write the SR functions as a function of angle and height in order to find similar expressions for Eq.~\ref{averageSurface1}-\ref{averageSurface2}. It is similarly expected for other boundary shapes that there is a larger chance to find an analytical expression with finite domain distributions instead of a bivariate normal distribution.}

\begin{figure}[tb]\centering
\includegraphics[width=0.4\linewidth]{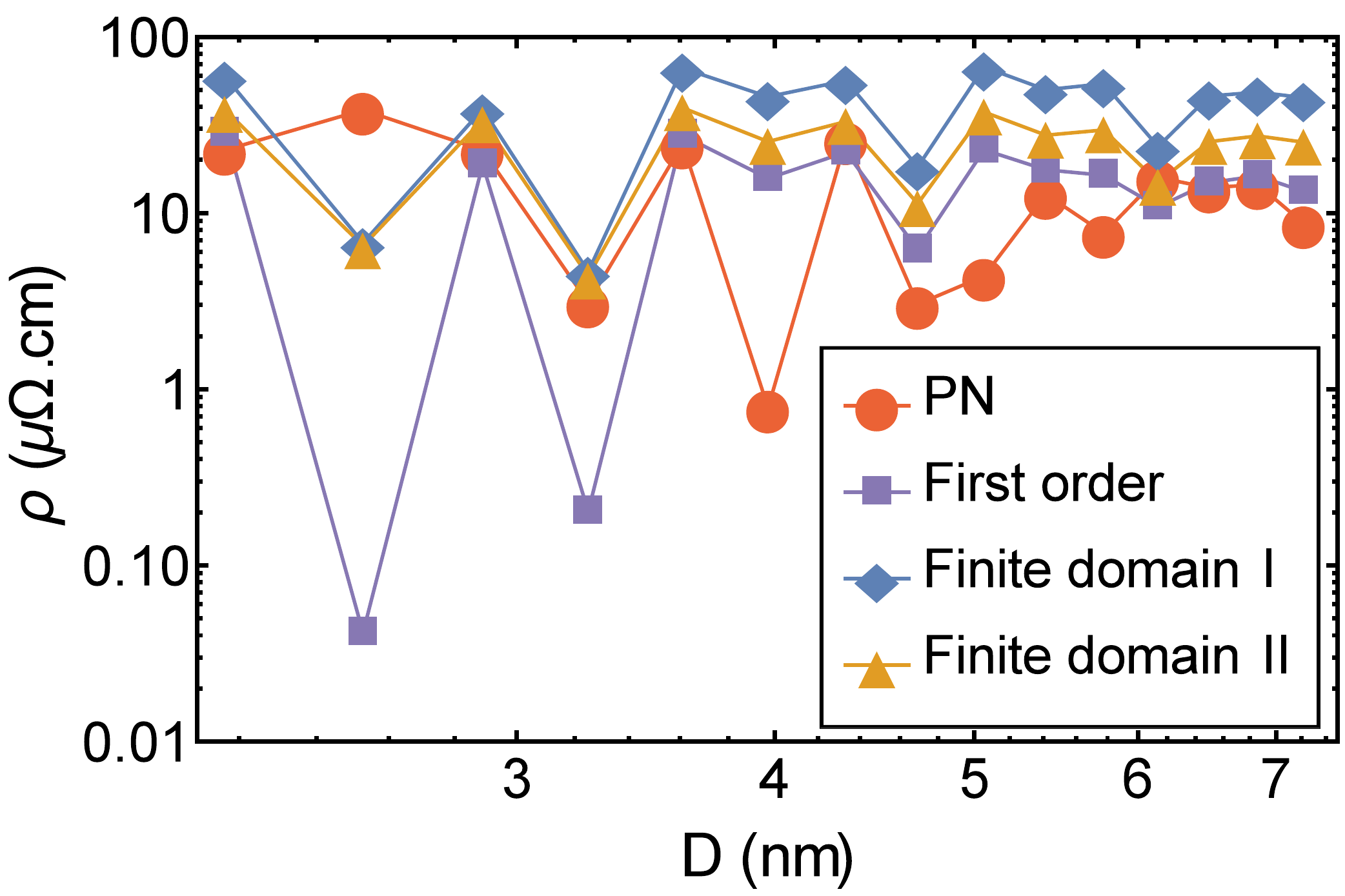}
\caption{The resistivity for copper nanowires with square cross section of sides ranging from 2 to 7~nm ($D \equiv L_x = L_y$, $\Delta = 2a_\textnormal{\tiny Cu} \approx 0.7$~nm, $\Lambda = 5a_\textnormal{\tiny Cu} \approx 1.75$~nm) are shown here, using the different methods explained in section \ref{sectionSRS}: Prange-Nee model (blue circles), first order model from Eq.~\ref{matrixElementInfiniteWell} (yellow squares), finite domain model I (green diamonds) and finite domain model II (red triangles).}
\label{condComparison}
\end{figure}

\begin{figure}[tb]\centering
\includegraphics[width=0.4\columnwidth]{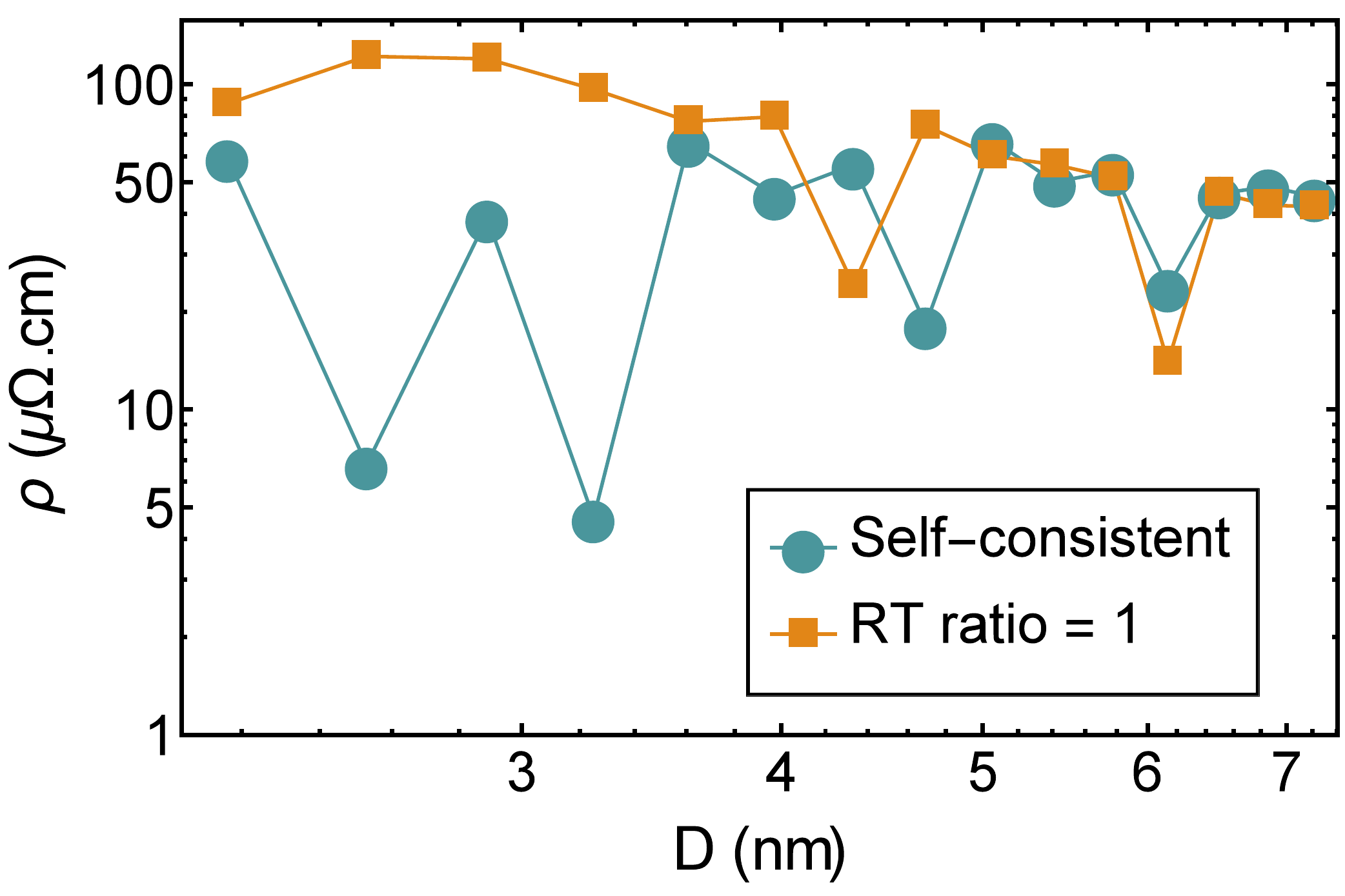}
\caption{The resistivity for copper nanowires with square cross section of sides ranging from 2 to 7~nm ($D \equiv L_x = L_y$) are shown here, using the self-consistent and approximate solution, substituting the relaxation time ratio on the RHS of Eq.~\ref{RTA} by 1, of the BTE with SR matrix elements obtained from finite domain model I ($\Delta = 2a_\textnormal{\tiny Cu} \approx 0.7$~nm, $\Lambda = 5a_\textnormal{\tiny Cu} \approx 1.75$~nm).}
\label{PrangeNeeRes}
\end{figure}

\begin{figure*}[tb]\begin{center}
\subfigure[\ Subbands $D\approx 2.15$~nm]{
\includegraphics[width=0.35\linewidth]{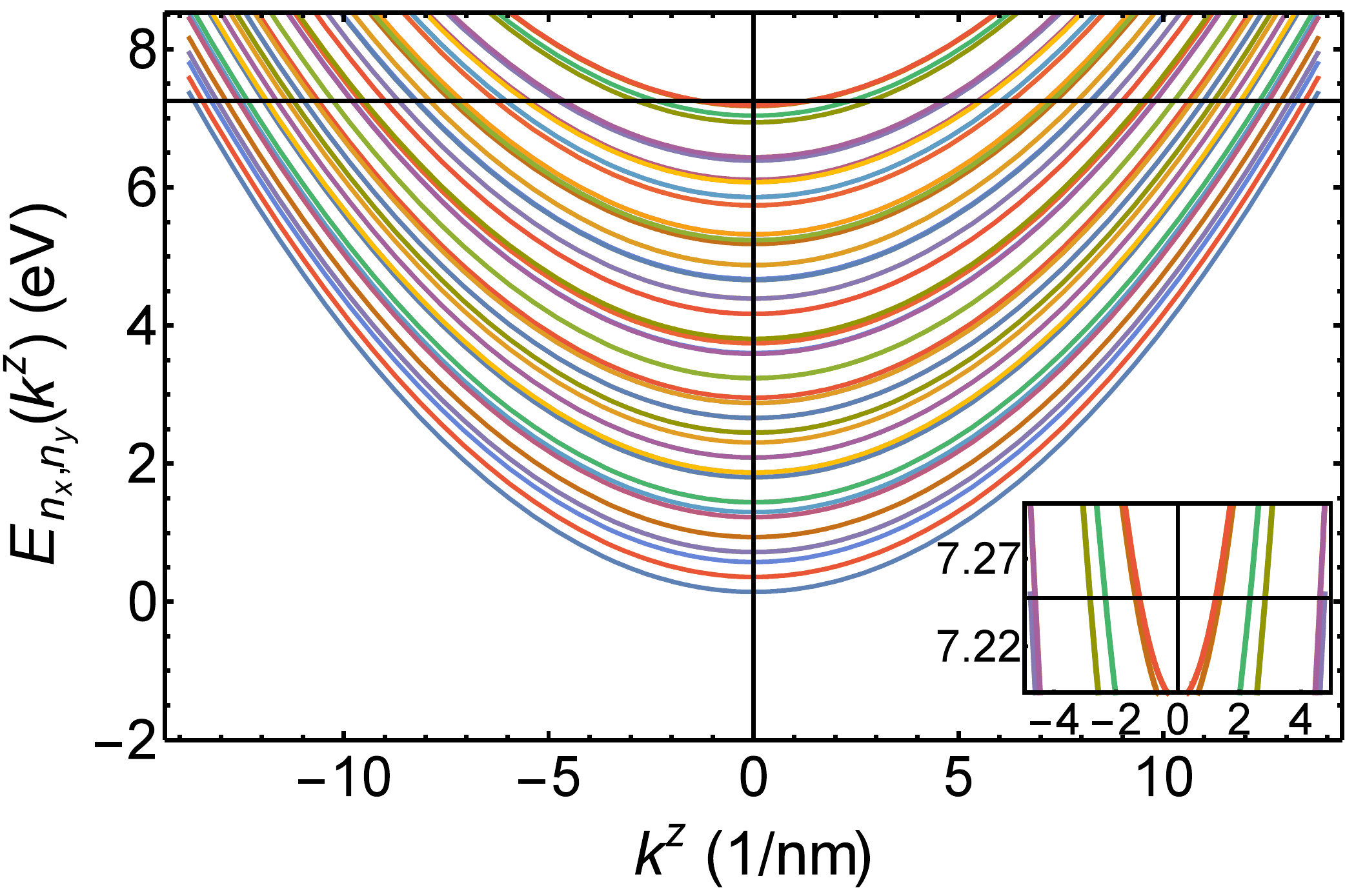}}
\subfigure[\ Subbands $D\approx 3.25$~nm]{
\includegraphics[width=0.35\linewidth]{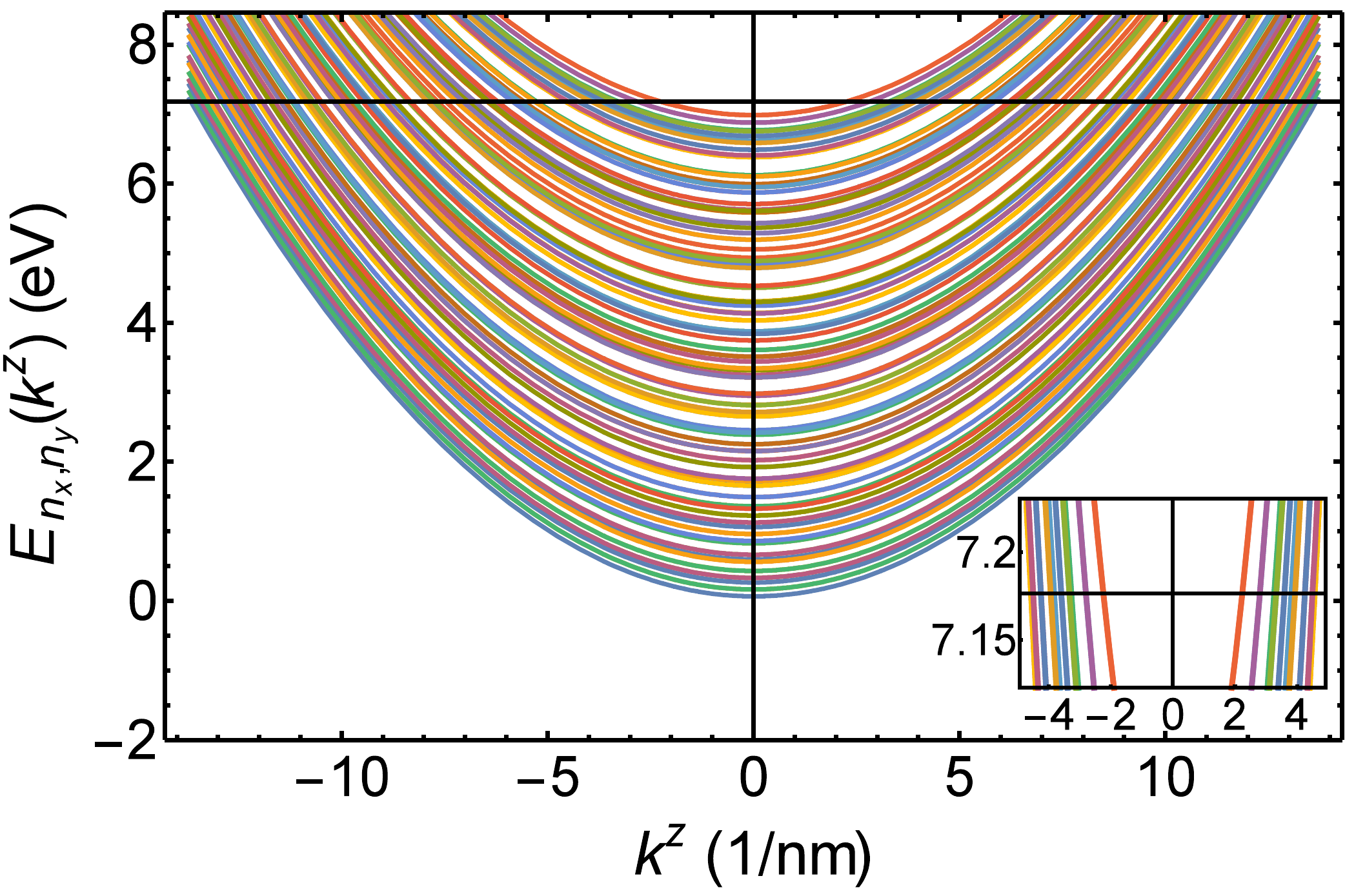}}
\subfigure[\ Relaxation times $D\approx 2.15$~nm]{
\includegraphics[width=0.35\linewidth]{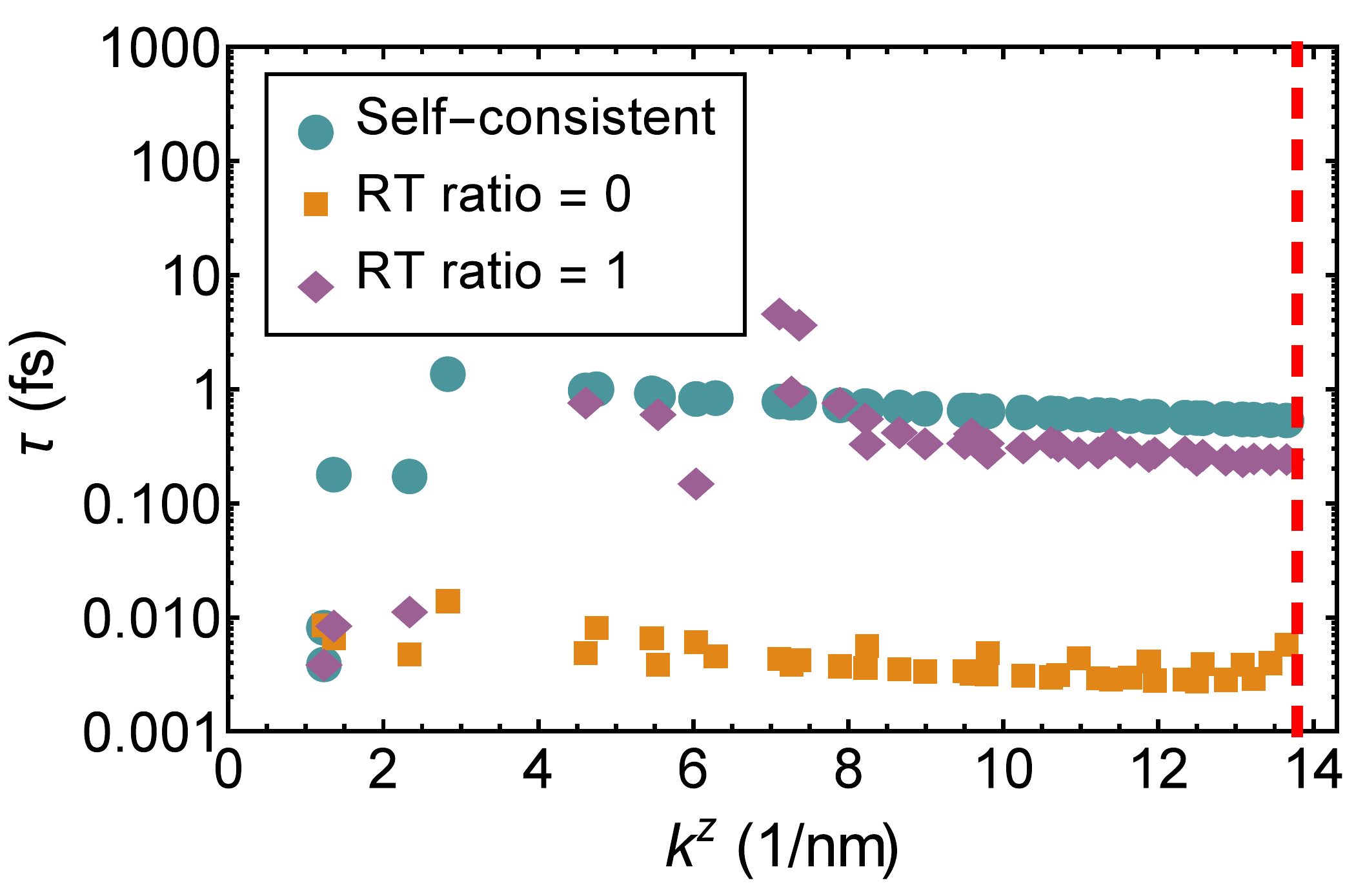}}
\subfigure[\ Relaxation times $D\approx 3.25$~nm]{
\includegraphics[width=0.35\linewidth]{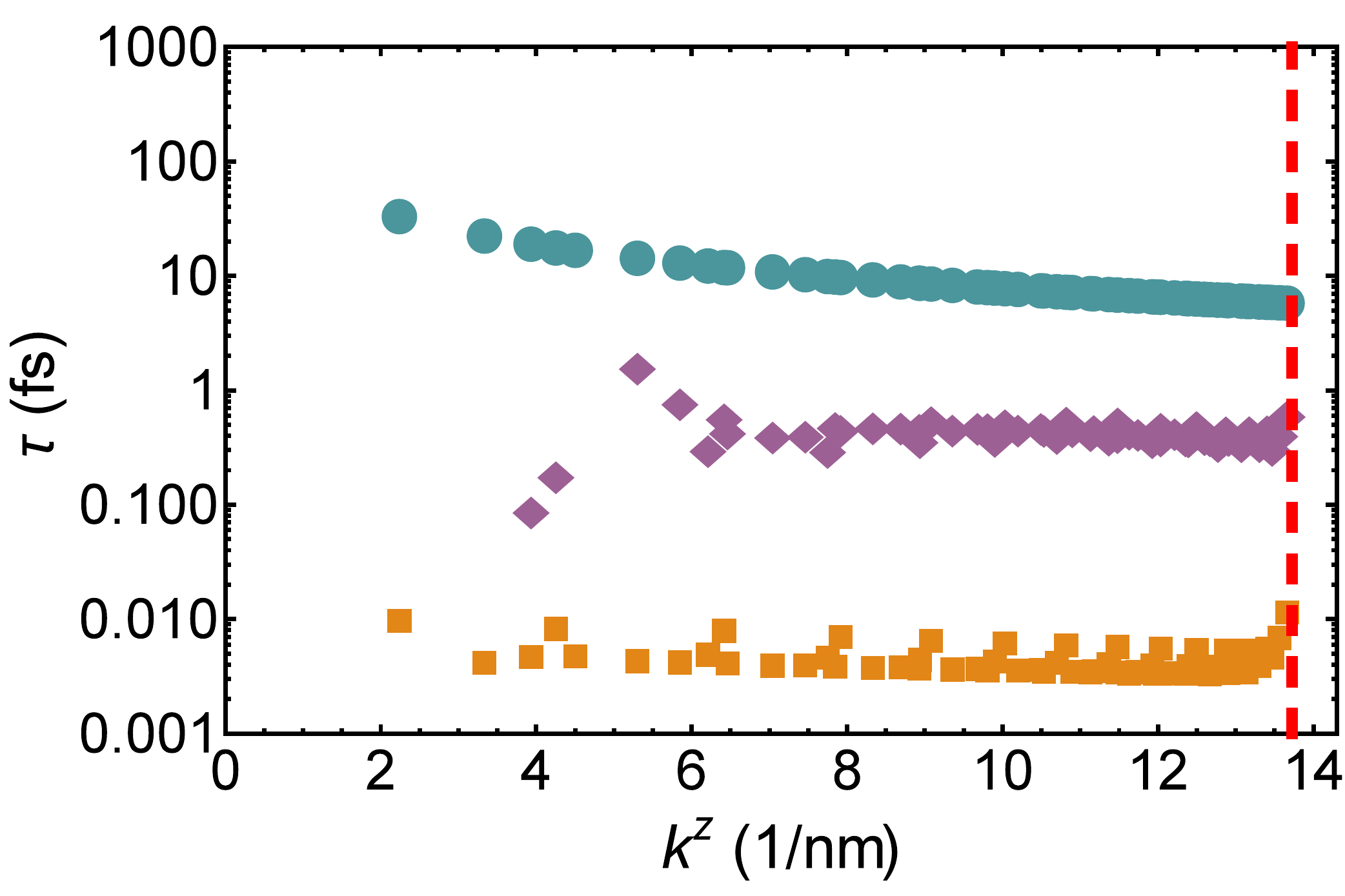}}
\end{center}
\caption{All the subbands that cross the Fermi level of a copper wire with $L_x = L_y$ equal to (a) 6 $a_\textnormal{\tiny Cu} \approx 2.15$~nm, (b) 9 $a_\textnormal{\tiny Cu} \approx 3.25$~nm are shown as a function of $k^z$. The self-consistent (green circles) and approximate (relaxation time ratio = 0: yellow squares, relaxation time ratio = 1: purple diamonds) relaxation time solutions of Eq.~\ref{RTA} with SR matrix elements, obtained with finite domain model I ($\Delta = 2a_\textnormal{\tiny Cu} \approx 0.7$~nm, $\Lambda = 5a_\textnormal{\tiny Cu} \approx 1.75$~nm), are shown for every Fermi level state with positive $k^z$ in femtosecond log scale in (c-d). The red dashed line is showing the Fermi wave vector $k_\textnormal{\tiny F} \equiv \sqrt{2m_e E_\textnormal{\tiny F}}/\hbar$.}
\label{PrangeNeeGrid}
\end{figure*}

The resistivities, obtained with the different methods and the self-consistent solution of the BTE, are compared in Fig.~\ref{condComparison}. In contrast to the one-band toy model, finite domain model I now differs most from the other methods and Prange-Nee works better, although it does not follow the same trend as all other methods. The first order approximation and finite domain model II now seem to be closest in agreement, with only small difference in resistivity value for all diameters. The overall trend for decreasing diameters is an increase of the resistivity (not with the Prange-Nee model results), but the results show a very low resistivity value for the wire with sides between 3 and 3.5~nm for all methods. Comparing the very low resistivity values obtained with the different methods, we observe quite big differences. This can be expected because the important states for back-scattering have quite high wave vectors in the confinement directions and the different methods compute the wave function overlap due to roughness quite differently for these states. The resistivity drop appears to be related to the size of the minimal momentum gap between the positive and negative $k^z$ state with lowest absolute value. When this gap exceeds the critical momentum gap (Eq.~\ref{critMomentumGapPN}), back-scattering is highly suppressed and only scattering to other states without substantial loss of forward momentum or current takes place. This can be seen in Fig.~\ref{PrangeNeeGrid}, where the band structure and relaxation time of the two lowest diameter data points of Fig.~\ref{PrangeNeeRes} are shown. The momentum gap of the larger diameter is approximately $\Delta k^z_\textnormal{crit.}$, which increases the lifetime of the low $k^z$ states drastically.

\setlength{\unitlength}{1.2cm}

Another interesting aspect of the resistivity drop is that it is only visible in the self-consistent solution of Eq.~\ref{RTA}. This solution is shown in Fig.~\ref{PrangeNeeRes} together with the approximated solution by putting the relaxation time ratio on the right-hand-side equal to 1, assuming equal lifetimes for the initial and final states.\cite{moors2014resistivity} The resistivity drop is only apparent in the self-consistent solution of the BTE because it is the only solution that solves the coupled system of equations, determining the scattering lifetimes correctly. Two data points of Fig.~\ref{PrangeNeeRes} are analyzed in more detail in Fig.~\ref{PrangeNeeGrid}. From that figure we can estimate the critical gap $\Delta k^z_\textnormal{crit.}$ and it appears to agree with $\Delta k^z_\textnormal{crit.} \approx 2\textnormal{ nm}^{-1}$, obtained from Eq.~\ref{critMomentumGapPN}. When the incoming scattering is neglected in Eq.~\ref{RTA}, by putting the relaxation time ratio equal to zero, the relaxation times are largely underestimated and the predicted resistivity is largely overestimated.

Finally, suppose that the momentum gap is around 1 or 2~nm${}^{-1}$, typically the case for nanowires with width and height up to 10~nm. Then the critical correlation length $\Lambda_\textnormal{crit.}$ will be of the order of 1~nm. If the SR correlation length is substantially larger than $\Lambda_\textnormal{crit.}$, the scattering matrix element will be exponentially suppressed by Eq.~\ref{deltaKzCrit} (or a similar expression for finite barrier heights) and the resistivity contribution becomes very small, even if the SR standard deviation is quite large. This is in agreement with results reported by Leunissen et al.\cite{leunissen2006impact} who found that the influence of roughness added to a wire has no impact. The critical correlation length is on an atomic scale, rather than the scale of the wire dimensions or the scale related to the etch process (typically larger than the atomic scale). Roughness properties on atomic scale have already been considered in semiconductor device modeling,\cite{mazzoni1999surface,esseni2004modeling,jin2007modeling,lizzit2014new} and experimentally verified for those types of boundaries with AFM and TEM.\cite{goodnick1985surface,yamanaka1996correlation,pirovano2000correlation,bonno2008effect}. The SR on atomic scale appears to be crucial for metallic nanowire modeling as well and it is very unlikely to be much dependent on wire width or height. Therefore the SR standard deviation and correlation length were considered to be independent of the wire dimensions in the simulations presented above.

\section{Conclusion}
\label{sectionConclusion}
The investigation of SR scattering, partially responsible for an increased resistivity in narrow metallic nanowires, requires a good model for comparison with experimental data. We propose a model based on Ando's model to model SR scattering exploiting the statistical features of roughness, rather than more empirical approaches like the Fuchs-Sondheimer model. With this model we can compare experimental data for metallic nanowires both on the level of resistivity measurements and detailed SR properties. This allows for proper testing of the SR model as well as improved resistivity predictions and simulations.

A crucial observation is that the BTE under the relaxation time approximation should be solved self-consistently to obtain the relaxation times and resulting resistivity correctly for SR scattering. While this method was already proposed in Moors et al.,\cite{moors2014resistivity} the Prange-Nee approximation was always used before to obtain the SR matrix elements that are based on Ando's model. We have now shown that the Prange-Nee approximation has serious shortcomings by neglecting the wave function oscillations and considering an infinite barrier limit. Because the wave functions can be highly oscillating, an alternative approach has been presented here under two variants to obtain Ando's matrix elements. The two variants are consistent with Gaussian SR autocorrelation functions and allow for a fast and accurate evaluation of the SR matrix elements with any roughness size. The statistical properties of surface roughness are captured in both cases by an appropriate distribution function on a finite domain. This new approach has also been compared with a recent proposal that also gives the SR matrix element without assuming small roughness sizes, but requires numerical integration when a bivariate normal distribution function is invoked.

Comparing all the different methods, the Prange-Nee model surprisingly works quite well for resistivity estimation of metallic nanowires with sides of a few nanometers. It clearly gives wrong results in the single subband toy model due to a large difference between the finite and infinite potential well solutions. In general the first order model works much better, even though the oscillations of the wave functions normal to the boundary plane are completely neglected. The newly introduced methods are in good agreement for the resistivity scaling trend and are faster than the method based on a bivariate normal distribution function by two orders of magnitude.

An important observation in this work is a resistivity drop for certain diameters, confirmed by all the self-consistent BTE solution with all the different methods. The crucial parameter is the wave vector gap between the Fermi level states with negative and positive wave numbers. An exponential suppression of current-loss inducing scattering occurs when this gap exceeds a critical inverse length scale that is found to be proportional to the inverse of the SR correlation length for all the simulations. This allows for a conductivity enhancement by engineering the subbands in such a way that this gap is maximized, a proposal that was already made for semiconductor nanowires by Sakaki in the 80's.\cite{sakaki1980scattering,sakaki1986physical}

\begin{acknowledgments}
We would like to thank Prof. Karel Neto\v{c}n\'{y}, Prof. Christian Maes and Zsolt T\H okei for useful discussions.
\end{acknowledgments}

\bibliography{paperSurfaceRoughnessPrePrintBibTeX}{}

\begin{thebibliography}{31}%
\makeatletter
\providecommand \@ifxundefined [1]{%
 \@ifx{#1\undefined}
}%
\providecommand \@ifnum [1]{%
 \ifnum #1\expandafter \@firstoftwo
 \else \expandafter \@secondoftwo
 \fi
}%
\providecommand \@ifx [1]{%
 \ifx #1\expandafter \@firstoftwo
 \else \expandafter \@secondoftwo
 \fi
}%
\providecommand \natexlab [1]{#1}%
\providecommand \enquote  [1]{``#1''}%
\providecommand \bibnamefont  [1]{#1}%
\providecommand \bibfnamefont [1]{#1}%
\providecommand \citenamefont [1]{#1}%
\providecommand \href@noop [0]{\@secondoftwo}%
\providecommand \href [0]{\begingroup \@sanitize@url \@href}%
\providecommand \@href[1]{\@@startlink{#1}\@@href}%
\providecommand \@@href[1]{\endgroup#1\@@endlink}%
\providecommand \@sanitize@url [0]{\catcode `\\12\catcode `\$12\catcode
  `\&12\catcode `\#12\catcode `\^12\catcode `\_12\catcode `\%12\relax}%
\providecommand \@@startlink[1]{}%
\providecommand \@@endlink[0]{}%
\providecommand \url  [0]{\begingroup\@sanitize@url \@url }%
\providecommand \@url [1]{\endgroup\@href {#1}{\urlprefix }}%
\providecommand \urlprefix  [0]{URL }%
\providecommand \Eprint [0]{\href }%
\providecommand \doibase [0]{http://dx.doi.org/}%
\providecommand \selectlanguage [0]{\@gobble}%
\providecommand \bibinfo  [0]{\@secondoftwo}%
\providecommand \bibfield  [0]{\@secondoftwo}%
\providecommand \translation [1]{[#1]}%
\providecommand \BibitemOpen [0]{}%
\providecommand \bibitemStop [0]{}%
\providecommand \bibitemNoStop [0]{.\EOS\space}%
\providecommand \EOS [0]{\spacefactor3000\relax}%
\providecommand \BibitemShut  [1]{\csname bibitem#1\endcsname}%
\let\auto@bib@innerbib\@empty
\bibitem [{\citenamefont {Fuchs}\ \emph {et~al.}(1938)\citenamefont {Fuchs}
  \emph {et~al.}}]{fuchs1938conductivity}%
  \BibitemOpen
  \bibfield  {author} {\bibinfo {author} {\bibfnamefont {K.}~\bibnamefont
  {Fuchs}} \emph {et~al.},\ }in\ \href@noop {} {\emph {\bibinfo {booktitle}
  {Proceedings of Cambridge Philosophical Society}}},\ Vol.~\bibinfo {volume}
  {34}\ (\bibinfo {organization} {Cambridge Univ. Press},\ \bibinfo {year}
  {1938})\ p.\ \bibinfo {pages} {100}\BibitemShut {NoStop}%
\bibitem [{\citenamefont {Sondheimer}(1952)}]{sondheimer1952mean}%
  \BibitemOpen
  \bibfield  {author} {\bibinfo {author} {\bibfnamefont {E.~H.}\ \bibnamefont
  {Sondheimer}},\ }\href@noop {} {\bibfield  {journal} {\bibinfo  {journal}
  {Advances in Physics}\ }\textbf {\bibinfo {volume} {1}},\ \bibinfo {pages}
  {1} (\bibinfo {year} {1952})}\BibitemShut {NoStop}%
\bibitem [{\citenamefont {Durkan}\ and\ \citenamefont
  {Welland}(2000)}]{durkan2000size}%
  \BibitemOpen
  \bibfield  {author} {\bibinfo {author} {\bibfnamefont {C.}~\bibnamefont
  {Durkan}}\ and\ \bibinfo {author} {\bibfnamefont {M.~E.}\ \bibnamefont
  {Welland}},\ }\href@noop {} {\bibfield  {journal} {\bibinfo  {journal}
  {Physical Review B}\ }\textbf {\bibinfo {volume} {61}},\ \bibinfo {pages}
  {14215} (\bibinfo {year} {2000})}\BibitemShut {NoStop}%
\bibitem [{\citenamefont {Steinh{\"o}gl}\ \emph {et~al.}(2002)\citenamefont
  {Steinh{\"o}gl}, \citenamefont {Schindler}, \citenamefont {Steinlesberger},\
  and\ \citenamefont {Engelhardt}}]{steinhogl2002size}%
  \BibitemOpen
  \bibfield  {author} {\bibinfo {author} {\bibfnamefont {W.}~\bibnamefont
  {Steinh{\"o}gl}}, \bibinfo {author} {\bibfnamefont {G.}~\bibnamefont
  {Schindler}}, \bibinfo {author} {\bibfnamefont {G.}~\bibnamefont
  {Steinlesberger}}, \ and\ \bibinfo {author} {\bibfnamefont {M.}~\bibnamefont
  {Engelhardt}},\ }\href@noop {} {\bibfield  {journal} {\bibinfo  {journal}
  {Physical Review B}\ }\textbf {\bibinfo {volume} {66}},\ \bibinfo {pages}
  {075414} (\bibinfo {year} {2002})}\BibitemShut {NoStop}%
\bibitem [{\citenamefont {Steinlesberger}\ \emph {et~al.}(2002)\citenamefont
  {Steinlesberger}, \citenamefont {Engelhardt}, \citenamefont {Schindler},
  \citenamefont {Steinh{\"o}gl}, \citenamefont {Von~Glasow}, \citenamefont
  {Mosig},\ and\ \citenamefont {Bertagnolli}}]{steinlesberger2002electrical}%
  \BibitemOpen
  \bibfield  {author} {\bibinfo {author} {\bibfnamefont {G.}~\bibnamefont
  {Steinlesberger}}, \bibinfo {author} {\bibfnamefont {M.}~\bibnamefont
  {Engelhardt}}, \bibinfo {author} {\bibfnamefont {G.}~\bibnamefont
  {Schindler}}, \bibinfo {author} {\bibfnamefont {W.}~\bibnamefont
  {Steinh{\"o}gl}}, \bibinfo {author} {\bibfnamefont {A.}~\bibnamefont
  {Von~Glasow}}, \bibinfo {author} {\bibfnamefont {K.}~\bibnamefont {Mosig}}, \
  and\ \bibinfo {author} {\bibfnamefont {E.}~\bibnamefont {Bertagnolli}},\
  }\href@noop {} {\bibfield  {journal} {\bibinfo  {journal} {Microelectronic
  Engineering}\ }\textbf {\bibinfo {volume} {64}},\ \bibinfo {pages} {409}
  (\bibinfo {year} {2002})}\BibitemShut {NoStop}%
\bibitem [{\citenamefont {Guillaumond}\ \emph {et~al.}(2003)\citenamefont
  {Guillaumond}, \citenamefont {Arnaud}, \citenamefont {Mourier}, \citenamefont
  {Fayolle}, \citenamefont {Pesci},\ and\ \citenamefont
  {Reimbold}}]{guillaumond2003analysis}%
  \BibitemOpen
  \bibfield  {author} {\bibinfo {author} {\bibfnamefont {J.~F.}\ \bibnamefont
  {Guillaumond}}, \bibinfo {author} {\bibfnamefont {L.}~\bibnamefont {Arnaud}},
  \bibinfo {author} {\bibfnamefont {T.}~\bibnamefont {Mourier}}, \bibinfo
  {author} {\bibfnamefont {M.}~\bibnamefont {Fayolle}}, \bibinfo {author}
  {\bibfnamefont {O.}~\bibnamefont {Pesci}}, \ and\ \bibinfo {author}
  {\bibfnamefont {G.}~\bibnamefont {Reimbold}},\ }in\ \href@noop {} {\emph
  {\bibinfo {booktitle} {Interconnect Technology Conference, 2003. Proceedings
  of the IEEE 2003 International}}}\ (\bibinfo {organization} {IEEE},\ \bibinfo
  {year} {2003})\ pp.\ \bibinfo {pages} {132--134}\BibitemShut {NoStop}%
\bibitem [{\citenamefont {Wu}\ \emph {et~al.}(2004)\citenamefont {Wu},
  \citenamefont {Brongersma}, \citenamefont {Van~Hove},\ and\ \citenamefont
  {Maex}}]{wu2004influence}%
  \BibitemOpen
  \bibfield  {author} {\bibinfo {author} {\bibfnamefont {W.}~\bibnamefont
  {Wu}}, \bibinfo {author} {\bibfnamefont {S.~H.}\ \bibnamefont {Brongersma}},
  \bibinfo {author} {\bibfnamefont {M.}~\bibnamefont {Van~Hove}}, \ and\
  \bibinfo {author} {\bibfnamefont {K.}~\bibnamefont {Maex}},\ }\href@noop {}
  {\bibfield  {journal} {\bibinfo  {journal} {Applied Physics Letters}\
  }\textbf {\bibinfo {volume} {84}},\ \bibinfo {pages} {2838} (\bibinfo {year}
  {2004})}\BibitemShut {NoStop}%
\bibitem [{\citenamefont {Steinh{\"o}gl}\ \emph {et~al.}(2005)\citenamefont
  {Steinh{\"o}gl}, \citenamefont {Schindler}, \citenamefont {Steinlesberger},
  \citenamefont {Traving},\ and\ \citenamefont
  {Engelhardt}}]{steinhogl2004comprehensive}%
  \BibitemOpen
  \bibfield  {author} {\bibinfo {author} {\bibfnamefont {W.}~\bibnamefont
  {Steinh{\"o}gl}}, \bibinfo {author} {\bibfnamefont {G.}~\bibnamefont
  {Schindler}}, \bibinfo {author} {\bibfnamefont {G.}~\bibnamefont
  {Steinlesberger}}, \bibinfo {author} {\bibfnamefont {M.}~\bibnamefont
  {Traving}}, \ and\ \bibinfo {author} {\bibfnamefont {M.}~\bibnamefont
  {Engelhardt}},\ }\href@noop {} {\bibfield  {journal} {\bibinfo  {journal}
  {Journal of Applied Physics}\ }\textbf {\bibinfo {volume} {97}},\ \bibinfo
  {pages} {023706} (\bibinfo {year} {2005})}\BibitemShut {NoStop}%
\bibitem [{\citenamefont {Zhang}\ \emph {et~al.}(2007)\citenamefont {Zhang},
  \citenamefont {Brongersma}, \citenamefont {Li}, \citenamefont {Li},
  \citenamefont {Richard},\ and\ \citenamefont {Maex}}]{zhang2007analysis}%
  \BibitemOpen
  \bibfield  {author} {\bibinfo {author} {\bibfnamefont {W.}~\bibnamefont
  {Zhang}}, \bibinfo {author} {\bibfnamefont {S.~H.}\ \bibnamefont
  {Brongersma}}, \bibinfo {author} {\bibfnamefont {Z.}~\bibnamefont {Li}},
  \bibinfo {author} {\bibfnamefont {D.}~\bibnamefont {Li}}, \bibinfo {author}
  {\bibfnamefont {O.}~\bibnamefont {Richard}}, \ and\ \bibinfo {author}
  {\bibfnamefont {K.}~\bibnamefont {Maex}},\ }\href@noop {} {\bibfield
  {journal} {\bibinfo  {journal} {Journal of Applied Physics}\ }\textbf
  {\bibinfo {volume} {101}},\ \bibinfo {pages} {063703} (\bibinfo {year}
  {2007})}\BibitemShut {NoStop}%
\bibitem [{\citenamefont {Josell}, \citenamefont {Brongersma},\ and\
  \citenamefont {Tokei}(2009)}]{josell2009size}%
  \BibitemOpen
  \bibfield  {author} {\bibinfo {author} {\bibfnamefont {D.}~\bibnamefont
  {Josell}}, \bibinfo {author} {\bibfnamefont {S.~H.}\ \bibnamefont
  {Brongersma}}, \ and\ \bibinfo {author} {\bibfnamefont {Z.}~\bibnamefont
  {Tokei}},\ }\href@noop {} {\bibfield  {journal} {\bibinfo  {journal} {Annual
  Review of Materials Research}\ }\textbf {\bibinfo {volume} {39}},\ \bibinfo
  {pages} {231} (\bibinfo {year} {2009})}\BibitemShut {NoStop}%
\bibitem [{\citenamefont {Graham}\ \emph {et~al.}(2010)\citenamefont {Graham},
  \citenamefont {Alers}, \citenamefont {Mountsier}, \citenamefont {Shamma},
  \citenamefont {Dhuey}, \citenamefont {Cabrini}, \citenamefont {Geiss},
  \citenamefont {Read},\ and\ \citenamefont {Peddeti}}]{graham2010resistivity}%
  \BibitemOpen
  \bibfield  {author} {\bibinfo {author} {\bibfnamefont {R.~L.}\ \bibnamefont
  {Graham}}, \bibinfo {author} {\bibfnamefont {G.~B.}\ \bibnamefont {Alers}},
  \bibinfo {author} {\bibfnamefont {T.}~\bibnamefont {Mountsier}}, \bibinfo
  {author} {\bibfnamefont {N.}~\bibnamefont {Shamma}}, \bibinfo {author}
  {\bibfnamefont {S.}~\bibnamefont {Dhuey}}, \bibinfo {author} {\bibfnamefont
  {S.}~\bibnamefont {Cabrini}}, \bibinfo {author} {\bibfnamefont {R.~H.}\
  \bibnamefont {Geiss}}, \bibinfo {author} {\bibfnamefont {D.~T.}\ \bibnamefont
  {Read}}, \ and\ \bibinfo {author} {\bibfnamefont {S.}~\bibnamefont
  {Peddeti}},\ }\href@noop {} {\bibfield  {journal} {\bibinfo  {journal}
  {Applied Physics Letters}\ }\textbf {\bibinfo {volume} {96}},\ \bibinfo
  {pages} {042116} (\bibinfo {year} {2010})}\BibitemShut {NoStop}%
\bibitem [{\citenamefont {Chawla}\ \emph {et~al.}(2011)\citenamefont {Chawla},
  \citenamefont {Gstrein}, \citenamefont {O’Brien}, \citenamefont {Clarke},\
  and\ \citenamefont {Gall}}]{chawla2011electron}%
  \BibitemOpen
  \bibfield  {author} {\bibinfo {author} {\bibfnamefont {J.~S.}\ \bibnamefont
  {Chawla}}, \bibinfo {author} {\bibfnamefont {F.}~\bibnamefont {Gstrein}},
  \bibinfo {author} {\bibfnamefont {K.~P.}\ \bibnamefont {O’Brien}}, \bibinfo
  {author} {\bibfnamefont {J.~S.}\ \bibnamefont {Clarke}}, \ and\ \bibinfo
  {author} {\bibfnamefont {D.}~\bibnamefont {Gall}},\ }\href@noop {} {\bibfield
   {journal} {\bibinfo  {journal} {Physical Review B}\ }\textbf {\bibinfo
  {volume} {84}},\ \bibinfo {pages} {235423} (\bibinfo {year}
  {2011})}\BibitemShut {NoStop}%
\bibitem [{\citenamefont {Mayadas}\ and\ \citenamefont
  {Shatzkes}(1970)}]{mayadas1970electrical}%
  \BibitemOpen
  \bibfield  {author} {\bibinfo {author} {\bibfnamefont {A.~F.}\ \bibnamefont
  {Mayadas}}\ and\ \bibinfo {author} {\bibfnamefont {M.}~\bibnamefont
  {Shatzkes}},\ }\href@noop {} {\bibfield  {journal} {\bibinfo  {journal}
  {Physical Review B}\ }\textbf {\bibinfo {volume} {1}},\ \bibinfo {pages}
  {1382} (\bibinfo {year} {1970})}\BibitemShut {NoStop}%
\bibitem [{\citenamefont {Ando}, \citenamefont {Fowler},\ and\ \citenamefont
  {Stern}(1982)}]{ando1982electronic}%
  \BibitemOpen
  \bibfield  {author} {\bibinfo {author} {\bibfnamefont {T.}~\bibnamefont
  {Ando}}, \bibinfo {author} {\bibfnamefont {A.~B.}\ \bibnamefont {Fowler}}, \
  and\ \bibinfo {author} {\bibfnamefont {F.}~\bibnamefont {Stern}},\
  }\href@noop {} {\bibfield  {journal} {\bibinfo  {journal} {Reviews of Modern
  Physics}\ }\textbf {\bibinfo {volume} {54}},\ \bibinfo {pages} {437}
  (\bibinfo {year} {1982})}\BibitemShut {NoStop}%
\bibitem [{\citenamefont {Prange}\ and\ \citenamefont
  {Nee}(1968)}]{prange1968quantum}%
  \BibitemOpen
  \bibfield  {author} {\bibinfo {author} {\bibfnamefont {R.}~\bibnamefont
  {Prange}}\ and\ \bibinfo {author} {\bibfnamefont {T.-W.}\ \bibnamefont
  {Nee}},\ }\href@noop {} {\bibfield  {journal} {\bibinfo  {journal} {Physical
  Review}\ }\textbf {\bibinfo {volume} {168}},\ \bibinfo {pages} {779}
  (\bibinfo {year} {1968})}\BibitemShut {NoStop}%
\bibitem [{\citenamefont {Moors}\ \emph {et~al.}(2014)\citenamefont {Moors},
  \citenamefont {Sor{\'e}e}, \citenamefont {T{\H{o}}kei},\ and\ \citenamefont
  {Magnus}}]{moors2014resistivity}%
  \BibitemOpen
  \bibfield  {author} {\bibinfo {author} {\bibfnamefont {K.}~\bibnamefont
  {Moors}}, \bibinfo {author} {\bibfnamefont {B.}~\bibnamefont {Sor{\'e}e}},
  \bibinfo {author} {\bibfnamefont {Z.}~\bibnamefont {T{\H{o}}kei}}, \ and\
  \bibinfo {author} {\bibfnamefont {W.}~\bibnamefont {Magnus}},\ }\href@noop {}
  {\bibfield  {journal} {\bibinfo  {journal} {Journal of Applied Physics}\
  }\textbf {\bibinfo {volume} {116}},\ \bibinfo {pages} {063714} (\bibinfo
  {year} {2014})}\BibitemShut {NoStop}%
\bibitem [{\citenamefont {Lizzit}\ \emph {et~al.}(2014)\citenamefont {Lizzit},
  \citenamefont {Esseni}, \citenamefont {Palestri},\ and\ \citenamefont
  {Selmi}}]{lizzit2014new}%
  \BibitemOpen
  \bibfield  {author} {\bibinfo {author} {\bibfnamefont {D.}~\bibnamefont
  {Lizzit}}, \bibinfo {author} {\bibfnamefont {D.}~\bibnamefont {Esseni}},
  \bibinfo {author} {\bibfnamefont {P.}~\bibnamefont {Palestri}}, \ and\
  \bibinfo {author} {\bibfnamefont {L.}~\bibnamefont {Selmi}},\ }\href@noop {}
  {\bibfield  {journal} {\bibinfo  {journal} {Journal of Applied Physics}\
  }\textbf {\bibinfo {volume} {116}},\ \bibinfo {pages} {223702} (\bibinfo
  {year} {2014})}\BibitemShut {NoStop}%
\bibitem [{Note1()}]{Note1}%
  \BibitemOpen
  \bibinfo {note} {We refer here to the energy difference between the highest
  occupied single-electron state in the ground state and the bottom of the
  potential well for the conduction electrons.}\BibitemShut {Stop}%
\bibitem [{\citenamefont {Sakaki}(1980)}]{sakaki1980scattering}%
  \BibitemOpen
  \bibfield  {author} {\bibinfo {author} {\bibfnamefont {H.}~\bibnamefont
  {Sakaki}},\ }\href@noop {} {\bibfield  {journal} {\bibinfo  {journal}
  {Japanese Journal of Applied Physics}\ }\textbf {\bibinfo {volume} {19}},\
  \bibinfo {pages} {L735} (\bibinfo {year} {1980})}\BibitemShut {NoStop}%
\bibitem [{\citenamefont {Sakaki}(1986)}]{sakaki1986physical}%
  \BibitemOpen
  \bibfield  {author} {\bibinfo {author} {\bibfnamefont {H.}~\bibnamefont
  {Sakaki}},\ }\href@noop {} {\bibfield  {journal} {\bibinfo  {journal}
  {Quantum Electronics, IEEE Journal of}\ }\textbf {\bibinfo {volume} {22}},\
  \bibinfo {pages} {1845} (\bibinfo {year} {1986})}\BibitemShut {NoStop}%
\bibitem [{\citenamefont {Motohisa}\ and\ \citenamefont
  {Sakaki}(1992)}]{motohisa1992interface}%
  \BibitemOpen
  \bibfield  {author} {\bibinfo {author} {\bibfnamefont {J.}~\bibnamefont
  {Motohisa}}\ and\ \bibinfo {author} {\bibfnamefont {H.}~\bibnamefont
  {Sakaki}},\ }\href@noop {} {\bibfield  {journal} {\bibinfo  {journal}
  {Applied physics letters}\ }\textbf {\bibinfo {volume} {60}},\ \bibinfo
  {pages} {1315} (\bibinfo {year} {1992})}\BibitemShut {NoStop}%
\bibitem [{\citenamefont {Kittel}, \citenamefont {McEuen},\ and\ \citenamefont
  {McEuen}(1976)}]{kittel1976introduction}%
  \BibitemOpen
  \bibfield  {author} {\bibinfo {author} {\bibfnamefont {C.}~\bibnamefont
  {Kittel}}, \bibinfo {author} {\bibfnamefont {P.}~\bibnamefont {McEuen}}, \
  and\ \bibinfo {author} {\bibfnamefont {P.}~\bibnamefont {McEuen}},\
  }\href@noop {} {\emph {\bibinfo {title} {Introduction to solid state
  physics}}},\ Vol.~\bibinfo {volume} {8}\ (\bibinfo  {publisher} {Wiley New
  York},\ \bibinfo {year} {1976})\BibitemShut {NoStop}%
\bibitem [{Note2()}]{Note2}%
  \BibitemOpen
  \bibinfo {note} {For cylindrical nanowires, it is better to start from
  circular potential well solutions and to write the SR functions as a function
  of angle and height in order to find similar expressions for Eq.~\ref
  {averageSurface1}-\ref {averageSurface2}. It is similarly expected for other
  boundary shapes that there is a larger chance to find an analytical
  expression with finite domain distributions instead of a bivariate normal
  distribution.}\BibitemShut {Stop}%
\bibitem [{\citenamefont {Leunissen}\ \emph {et~al.}(2006)\citenamefont
  {Leunissen}, \citenamefont {Zhang}, \citenamefont {Wu},\ and\ \citenamefont
  {Brongersma}}]{leunissen2006impact}%
  \BibitemOpen
  \bibfield  {author} {\bibinfo {author} {\bibfnamefont {L.}~\bibnamefont
  {Leunissen}}, \bibinfo {author} {\bibfnamefont {W.}~\bibnamefont {Zhang}},
  \bibinfo {author} {\bibfnamefont {W.}~\bibnamefont {Wu}}, \ and\ \bibinfo
  {author} {\bibfnamefont {S.}~\bibnamefont {Brongersma}},\ }\href@noop {}
  {\bibfield  {journal} {\bibinfo  {journal} {Journal of Vacuum Science \&
  Technology B}\ }\textbf {\bibinfo {volume} {24}},\ \bibinfo {pages} {1859}
  (\bibinfo {year} {2006})}\BibitemShut {NoStop}%
\bibitem [{\citenamefont {Mazzoni}\ \emph {et~al.}(1999)\citenamefont
  {Mazzoni}, \citenamefont {Lacaita}, \citenamefont {Perron},\ and\
  \citenamefont {Pirovano}}]{mazzoni1999surface}%
  \BibitemOpen
  \bibfield  {author} {\bibinfo {author} {\bibfnamefont {G.}~\bibnamefont
  {Mazzoni}}, \bibinfo {author} {\bibfnamefont {A.~L.}\ \bibnamefont
  {Lacaita}}, \bibinfo {author} {\bibfnamefont {L.~M.}\ \bibnamefont {Perron}},
  \ and\ \bibinfo {author} {\bibfnamefont {A.}~\bibnamefont {Pirovano}},\
  }\href@noop {} {\bibfield  {journal} {\bibinfo  {journal} {Electron Devices,
  IEEE Transactions on}\ }\textbf {\bibinfo {volume} {46}},\ \bibinfo {pages}
  {1423} (\bibinfo {year} {1999})}\BibitemShut {NoStop}%
\bibitem [{\citenamefont {Esseni}(2004)}]{esseni2004modeling}%
  \BibitemOpen
  \bibfield  {author} {\bibinfo {author} {\bibfnamefont {D.}~\bibnamefont
  {Esseni}},\ }\href@noop {} {\bibfield  {journal} {\bibinfo  {journal}
  {Electron Devices, IEEE Transactions on}\ }\textbf {\bibinfo {volume} {51}},\
  \bibinfo {pages} {394} (\bibinfo {year} {2004})}\BibitemShut {NoStop}%
\bibitem [{\citenamefont {Jin}, \citenamefont {Fischetti},\ and\ \citenamefont
  {Tang}(2007)}]{jin2007modeling}%
  \BibitemOpen
  \bibfield  {author} {\bibinfo {author} {\bibfnamefont {S.}~\bibnamefont
  {Jin}}, \bibinfo {author} {\bibfnamefont {M.~V.}\ \bibnamefont {Fischetti}},
  \ and\ \bibinfo {author} {\bibfnamefont {T.-W.}\ \bibnamefont {Tang}},\
  }\href@noop {} {\bibfield  {journal} {\bibinfo  {journal} {Electron Devices,
  IEEE Transactions on}\ }\textbf {\bibinfo {volume} {54}},\ \bibinfo {pages}
  {2191} (\bibinfo {year} {2007})}\BibitemShut {NoStop}%
\bibitem [{\citenamefont {Goodnick}\ \emph {et~al.}(1985)\citenamefont
  {Goodnick}, \citenamefont {Ferry}, \citenamefont {Wilmsen}, \citenamefont
  {Liliental}, \citenamefont {Fathy},\ and\ \citenamefont
  {Krivanek}}]{goodnick1985surface}%
  \BibitemOpen
  \bibfield  {author} {\bibinfo {author} {\bibfnamefont {S.}~\bibnamefont
  {Goodnick}}, \bibinfo {author} {\bibfnamefont {D.}~\bibnamefont {Ferry}},
  \bibinfo {author} {\bibfnamefont {C.}~\bibnamefont {Wilmsen}}, \bibinfo
  {author} {\bibfnamefont {Z.}~\bibnamefont {Liliental}}, \bibinfo {author}
  {\bibfnamefont {D.}~\bibnamefont {Fathy}}, \ and\ \bibinfo {author}
  {\bibfnamefont {O.}~\bibnamefont {Krivanek}},\ }\href@noop {} {\bibfield
  {journal} {\bibinfo  {journal} {Physical Review B}\ }\textbf {\bibinfo
  {volume} {32}},\ \bibinfo {pages} {8171} (\bibinfo {year}
  {1985})}\BibitemShut {NoStop}%
\bibitem [{\citenamefont {Yamanaka}\ \emph {et~al.}(1996)\citenamefont
  {Yamanaka}, \citenamefont {Fang}, \citenamefont {Lin}, \citenamefont
  {Snyder},\ and\ \citenamefont {Helms}}]{yamanaka1996correlation}%
  \BibitemOpen
  \bibfield  {author} {\bibinfo {author} {\bibfnamefont {T.}~\bibnamefont
  {Yamanaka}}, \bibinfo {author} {\bibfnamefont {S.~J.}\ \bibnamefont {Fang}},
  \bibinfo {author} {\bibfnamefont {H.-C.}\ \bibnamefont {Lin}}, \bibinfo
  {author} {\bibfnamefont {J.~P.}\ \bibnamefont {Snyder}}, \ and\ \bibinfo
  {author} {\bibfnamefont {C.~R.}\ \bibnamefont {Helms}},\ }\href@noop {}
  {\bibfield  {journal} {\bibinfo  {journal} {Electron Device Letters, IEEE}\
  }\textbf {\bibinfo {volume} {17}},\ \bibinfo {pages} {178} (\bibinfo {year}
  {1996})}\BibitemShut {NoStop}%
\bibitem [{\citenamefont {Pirovano}\ \emph {et~al.}(2000)\citenamefont
  {Pirovano}, \citenamefont {Lacaita}, \citenamefont {Ghidini},\ and\
  \citenamefont {Tallarida}}]{pirovano2000correlation}%
  \BibitemOpen
  \bibfield  {author} {\bibinfo {author} {\bibfnamefont {A.}~\bibnamefont
  {Pirovano}}, \bibinfo {author} {\bibfnamefont {A.}~\bibnamefont {Lacaita}},
  \bibinfo {author} {\bibfnamefont {G.}~\bibnamefont {Ghidini}}, \ and\
  \bibinfo {author} {\bibfnamefont {G.}~\bibnamefont {Tallarida}},\ }\href@noop
  {} {\bibfield  {journal} {\bibinfo  {journal} {Electron Device Letters,
  IEEE}\ }\textbf {\bibinfo {volume} {21}},\ \bibinfo {pages} {34} (\bibinfo
  {year} {2000})}\BibitemShut {NoStop}%
\bibitem [{\citenamefont {Bonno}\ \emph {et~al.}(2008)\citenamefont {Bonno},
  \citenamefont {Barraud}, \citenamefont {Mariolle},\ and\ \citenamefont
  {Andrieu}}]{bonno2008effect}%
  \BibitemOpen
  \bibfield  {author} {\bibinfo {author} {\bibfnamefont {O.}~\bibnamefont
  {Bonno}}, \bibinfo {author} {\bibfnamefont {S.}~\bibnamefont {Barraud}},
  \bibinfo {author} {\bibfnamefont {D.}~\bibnamefont {Mariolle}}, \ and\
  \bibinfo {author} {\bibfnamefont {F.}~\bibnamefont {Andrieu}},\ }\href@noop
  {} {\bibfield  {journal} {\bibinfo  {journal} {Journal of Applied Physics}\
  }\textbf {\bibinfo {volume} {103}},\ \bibinfo {pages} {063715} (\bibinfo
  {year} {2008})}\BibitemShut {NoStop}%
\end{thebibliography}%

\appendix

\section{Distribution functions on finite domain}
\label{sectionAppendix}
The expression for the average SR matrix element, using the distribution functions proposed in section \ref{sectionFin1} and section \ref{sectionFin2}, is quite involved and it is easy to loose track of the calculation. The full expressions for the matrix elements containing the four rough boundaries of a wire (with equal distribution functions and mutually uncorrelated) that were used in the simulations are given below for reference:
\begin{widetext}
\begin{align}
\left\langle \left| \langle i \mid V \mid f \rangle \right|^2 \right\rangle &= \textnormal{UNCORR}_{i,f} + \textnormal{CORR}^{\textnormal{I/II}}_{i,f}, \\ \notag
\textnormal{UNCORR}_{i,f} &\equiv \frac{4 \sin^2\left[ \lef k^z_i - k^z_f \rig L_z / 2 \right]}{L_z^2 \lef k^z_i - k^z_f \rig^2} \left| \int\limits_0^{L_y} \mkern-3mu \deriv y \; \psi_i\lef y \rig \psi_f\lef y \rig \lef \left\langle \textnormal{SI}_{i,f}^{x=0} \right\rangle + \left\langle \textnormal{SI}_{i,f}^{x=L_x} \right\rangle \rig \right. \\ \notag
& \qquad \qquad \qquad \qquad \qquad \qquad + \left. \int\limits_0^{L_x} \mkern-3mu \deriv x \; \psi_i\lef x \rig \psi_f\lef x \rig \lef \left\langle \textnormal{SI}_{i,f}^{y=0} \right\rangle + \left\langle \textnormal{SI}_{i,f}^{y=L_y} \right\rangle \rig \right|^2,
\end{align}
\begin{align*}
\textnormal{CORR}^{\textnormal{I}}_{i,f} &\equiv \frac{1}{L_z^2} \mkern-8mu \int\limits_{-L_z/2}^{+L_z/2} \mkern-12mu \deriv z \mkern-8mu \int\limits_{-L_z/2}^{+L_z/2} \mkern-12mu \deriv \tilde{z} \; e^{ -i (k^z_i - k^z_f) (z - z') - (z-z')^2/(\Lambda^2 / 2)} \\ \notag
& \quad \times \left\{ \int\limits_0^{L_y} \mkern-3mu \deriv y \mkern-3mu \int\limits_0^{L_y} \mkern-3mu \deriv \tilde{y} \; \psi_i (y) \psi_f (y) \psi_f (\tilde{y}) \psi_i (\tilde{y}) e^{- (y - \tilde{y})^2/(\Lambda^2/2)} \right. \\
& \qquad \qquad \times \frac{4}{3} \left[ \left| \left\langle \textnormal{sign}\lef S^{x=0} \rig \textnormal{SI}_{i,f}^{x=0} \right\rangle \right|^2 + \left| \left\langle \textnormal{sign}\lef S^{x=L_x} \rig \textnormal{SI}_{i,f}^{x=L_x} \right\rangle \right|^2 \right] \\ \notag
& \qquad \quad + \int\limits_0^{L_x} \mkern-3mu \deriv x \mkern-3mu \int\limits_0^{L_x} \mkern-3mu \deriv \tilde{x} \; \psi_i (x) \psi_f (x) \psi_f (\tilde{x}) \psi_i (\tilde{x}) e^{- (x - \tilde{x})^2/(\Lambda^2/2)} \\ \notag
& \qquad \qquad \times \left. \frac{4}{3} \left[ \left| \left\langle \textnormal{sign}\lef S^{y=0} \rig \textnormal{SI}_{i,f}^{y=0} \right\rangle \right|^2 + \left| \left\langle \textnormal{sign}\lef S^{y=L_y} \rig \textnormal{SI}_{i,f}^{y=L_y} \right\rangle \right|^2 \right] \right\},
\end{align*}
\begin{align*}
\textnormal{CORR}^{\textnormal{II}}_{i,f} &\equiv \frac{1}{L_z^2} \mkern-8mu \int\limits_{-L_z/2}^{+L_z/2} \mkern-12mu \deriv z \mkern-8mu \int\limits_{-L_z/2}^{+L_z/2} \mkern-12mu \deriv \tilde{z} \; e^{ -i (k^z_i - k^z_f) (z - \tilde{z}) - (z-\tilde{z})^2/(\Lambda^2/2)} \\ \notag
&\quad \times \left[ \int\limits_0^{L_y} \mkern-3mu \deriv y \mkern-3mu \int\limits_0^{L_y} \mkern-3mu \deriv \tilde{y} \; \psi_i (y) \psi_f (y) \psi_f (\tilde{y}) \psi_i (\tilde{y}) e^{- (y - \tilde{y})^2/(\Lambda^2 / 2)} \right. \\
&\qquad \quad \times \lef \left\langle \left| \textnormal{SI}_{i,f}^{x=0} \right|^2 \right\rangle - \left| \left\langle \textnormal{SI}_{i,f}^{x=0} \right\rangle \right|^2 + \left\langle \left| \textnormal{SI}_{i,f}^{x=L_x} \right|^2 \right\rangle - \left| \left\langle \textnormal{SI}_{i,f}^{x=L_x} \right\rangle \right|^2 \rig \\ \notag
&\qquad \quad + \int\limits_0^{L_x} \mkern-3mu \deriv x \mkern-3mu \int\limits_0^{L_x} \mkern-3mu \deriv \tilde{x} \; \psi_i (x) \psi_f (x) \psi_f (\tilde{x}) \psi_i (\tilde{x}) e^{- (x - \tilde{x})^2/(\Lambda^2/2)} \\ \notag
&\qquad \qquad \times \left. \lef \left\langle \left| \textnormal{SI}_{i,f}^{y=0} \right|^2 \right\rangle - \left| \left\langle \textnormal{SI}_{i,f}^{y=0} \right\rangle \right|^2 + \left\langle \left| \textnormal{SI}_{i,f}^{y=L_y} \right|^2 \right\rangle - \left| \left\langle \textnormal{SI}_{i,f}^{y=L_y} \right\rangle \right|^2 \rig \right].
\end{align*}
\end{widetext}
Note that even though the integration results are left implicit, every integral in the expressions can be written as an analytical expression.

\end{document}